\documentclass[twocolumn]{aastex631}

\usepackage[toc,page]{appendix}
\usepackage[T1]{fontenc}
\usepackage{textcomp} 
\usepackage{url}       
\usepackage{xspace}
\usepackage{amsmath}
\usepackage{rotating}
\usepackage{enumitem}
\usepackage{multirow}
\usepackage{tabularx}  
\usepackage{array}
\usepackage{hyperref}  
\hypersetup{linkcolor=blue, citecolor=blue, filecolor=cyan, urlcolor=blue}
\newcolumntype{C}{>{\centering\arraybackslash}X}

\graphicspath{{./}}

\shorttitle{The Sizes of LAEs at $3 \lesssim z < 7$}
\shortauthors{Song et al.}

\begin{document}

\title{The Size Evolution and the Size-Mass Relation of Lyman-Alpha Emitters across $3 \lesssim z < 7$ \\ 
as Observed by JWST}

\author[0009-0007-5833-3210]{Qi Song}
\affil{National Astronomical Observatories, Chinese Academy of Sciences, 20A Datun Road, Chaoyang District, Beijing 100101, China}
\affil{Key Laboratory of Optical Astronomy, National Astronomical Observatories, Chinese Academy of Sciences, 20A Datun Road, Chaoyang District, Beijing 100101, China}

\author[0009-0001-7105-2284]{F. S. Liu $^{\color{blue} \dagger}$}
\affil{National Astronomical Observatories, Chinese Academy of Sciences, 20A Datun Road, Chaoyang District, Beijing 100101, China}
\affil{Key Laboratory of Optical Astronomy, National Astronomical Observatories, Chinese Academy of Sciences, 20A Datun Road, Chaoyang District, Beijing 100101, China}
\affil{School of Astronomy and Space Science, University of Chinese Academy of Science, Beijing 100049, China}

\author[0000-0002-5043-2886]{Jian Ren $^{\color{blue} \dagger}$} 
\affil{National Astronomical Observatories, Chinese Academy of Sciences, 20A Datun Road, Chaoyang District, Beijing 100101, China}
\affil{Key Laboratory of Space Astronomy and Technology, National Astronomical Observatories, Chinese Academy of Sciences, 20A Datun Road, Chaoyang District, Beijing 100101, China}

\author{Pinsong Zhao}
\affil{Kavli Institute for Astronomy and Astrophysics, Peking University, Beijing 100871, China}
\affil{National Astronomical Observatories, Chinese Academy of Sciences, 20A Datun Road, Chaoyang District, Beijing 100101, China}

\author[0009-0001-5320-1450]{Qifan Cui}
\affil{Shanghai Key Lab for Astrophysics, Shanghai Normal University, Shanghai 200234, China}
\affil{National Astronomical Observatories, Chinese Academy of Sciences, 20A Datun Road, Chaoyang District, Beijing 100101, China}

\author[0000-0002-4882-1057]{Yubin Li}
\affil{National Astronomical Observatories, Chinese Academy of Sciences, 20A Datun Road, Chaoyang District, Beijing 100101, China}

\author[0000-0002-3443-0768]{Hao Mo}
\affil{National Astronomical Observatories, Chinese Academy of Sciences, 20A Datun Road, Chaoyang District, Beijing 100101, China}
\affil{Key Laboratory of Optical Astronomy, National Astronomical Observatories, Chinese Academy of Sciences, 20A Datun Road, Chaoyang District, Beijing 100101, China}
\affil{School of Astronomy and Space Science, University of Chinese Academy of Science, Beijing 100049, China}

\author{Yuchong Luo}
\affil{National Astronomical Observatories, Chinese Academy of Sciences, 20A Datun Road, Chaoyang District, Beijing 100101, China}

\author{Guanghuan Wang}
\affil{Purple Mountain Observatory, Chinese Academy of Sciences, 10 Yuanhua Road, Nanjing 210034, China}
\affil{National Astronomical Observatories, Chinese Academy of Sciences, 20A Datun Road, Chaoyang District, Beijing 100101, China}

\author{Nan Li $^{\color{blue} \dagger}$}
\affil{National Astronomical Observatories, Chinese Academy of Sciences, 20A Datun Road, Chaoyang District, Beijing 100101, China}
\affil{Key Laboratory of Space Astronomy and Technology, National Astronomical Observatories, Chinese Academy of Sciences, 20A Datun Road, Chaoyang District, Beijing 100101, China}
\affil{School of Astronomy and Space Science, University of Chinese Academy of Science, Beijing 100049, China}

\author{Hassen M. Yesuf}
\affil{Key Laboratory for Research in Galaxies and Cosmology, Shanghai Astronomical Observatory, Chinese Academy of Sciences, 80 Nandan Road, Shanghai 200030, China}

\author{Weichen Wang}
\affil{Dipartimento di Fisica G. Occhialini, Università degli Studi di Milano-Bicocca, Piazza della Scienza 3, I-20126 Milano, Italy}

\author{Xin Zhang}
\affil{National Astronomical Observatories, Chinese Academy of Sciences, 20A Datun Road, Chaoyang District, Beijing 100101, China}
\affil{Key Laboratory of Space Astronomy and Technology, National Astronomical Observatories, Chinese Academy of Sciences, 20A Datun Road, Chaoyang District, Beijing 100101, China}

\author{Xianmin Meng}
\affil{National Astronomical Observatories, Chinese Academy of Sciences, 20A Datun Road, Chaoyang District, Beijing 100101, China}
\affil{Key Laboratory of Space Astronomy and Technology, National Astronomical Observatories, Chinese Academy of Sciences, 20A Datun Road, Chaoyang District, Beijing 100101, China}

\author{Mingxiang Fu}
\affil{National Astronomical Observatories, Chinese Academy of Sciences, 20A Datun Road, Chaoyang District, Beijing 100101, China}
\affil{Key Laboratory of Space Astronomy and Technology, National Astronomical Observatories, Chinese Academy of Sciences, 20A Datun Road, Chaoyang District, Beijing 100101, China}
\affil{School of Astronomy and Space Science, University of Chinese Academy of Science, Beijing 100049, China}

\author{Bingqing Zhang}
\affil{National Astronomical Observatories, Chinese Academy of Sciences, 20A Datun Road, Chaoyang District, Beijing 100101, China}
\affil{Key Laboratory of Optical Astronomy, National Astronomical Observatories, Chinese Academy of Sciences, 20A Datun Road, Chaoyang District, Beijing 100101, China}

\author{Chenxiaoji Ling}
\affil{National Astronomical Observatories, Chinese Academy of Sciences, 20A Datun Road, Chaoyang District, Beijing 100101, China}
\affil{Key Laboratory of Space Astronomy and Technology, National Astronomical Observatories, Chinese Academy of Sciences, 20A Datun Road, Chaoyang District, Beijing 100101, China}

\begin{abstract}

Understanding the morphological structures of Lyman-alpha emitters (LAEs) is crucial 
for unveiling their formation pathways and the physical origins of Ly$\alpha$ emission. 
However, the evolution of their sizes and structural scaling relations remains debated. 
In this study, we analyze a large sample of 876 spectroscopically confirmed LAEs 
at $3 \lesssim z < 7$, selected from the MUSE, VANDELS, and CANDELSz7 surveys 
in the GOODS-S, UDS, and COSMOS fields. 
Utilizing James Webb Space Telescope (JWST) NIRCam imaging data, 
we measure their rest-frame UV and optical V-band effective radii ($R_{\rm e}$) 
through two-dimensional S\'{e}rsic profile fitting. 
Our results show that these LAEs are generally compact, with a median $R_{\rm e,UV}$ of 0.50$^{+0.30}_{-0.24}$ kpc and a median $R_{\rm e,V}$ of 0.57$^{+0.33}_{-0.24}$ kpc. 
The size evolution follows $R_{\rm e,UV} \propto (1 + z)^{-0.91 \pm 0.10}$ and 
$R_{\rm e,V} \propto (1 + z)^{-0.93 \pm 0.18}$, respectively. 
Their UV and optical sizes are statistically comparable, indicating negligible UV-to-optical color gradients. 
For the first time, we establish the rest-frame optical size-mass relation for LAEs at $z>3$, 
finding slopes comparable to typical star-forming galaxies (SFGs), 
but with slightly smaller sizes at a given stellar mass. 
These results provide important clues for understanding structural evolution of LAEs in the early universe.

\end{abstract}


\email{E-mail: fsliu@nao.cas.cn; renjian@nao.cas.cn}

\section{Introduction} \label{sec:intro}

Lyman-alpha emitters (LAEs) are a class of galaxies characterized by intense Lyman-alpha (Ly$\alpha$, $\lambda$ = 121.6 nm) emission lines. This prominent spectral feature serves not only as a key tracer of early star formation and galaxy assembly, but also provides critical insights into the ionization history of the intergalactic medium (IGM) during cosmic reionization, the gas dynamics between galaxies and their circumgalactic environments, and the connection between star-forming regions and dark matter halo growth \citep{Dijkstra2014PASA...31...40D, 2020ARA&A..58..617O}.

Following the seminal theoretical prediction of LAEs by \citet{Partridge10.1086/149079}, 
advances in narrowband (NB) imaging and spectroscopic surveys have enabled the discovery of increasingly large and well-characterized LAE samples \citep{Hu1996Natur.382..231H, Hu2010ApJ...725..394H, Ouchi2010ApJ...723..869O, Jiang2017ApJ...846..134J, Zheng2017ApJ...842L..22Z, Shibuya2018PASJ...70S..15S, Ning2020ApJ...903....4N, Ning2022ApJ...926..230N}. These expanding datasets have provided a critical empirical foundation for investigating the universal physical properties of LAEs across cosmic time.

As typical high-redshift star-forming galaxies, LAEs are characterized by relatively low stellar masses ($M_* \sim 10^8$--$10^9\ M_\odot$), young stellar ages ($\sim$ 1--100 Myr), star formation rates (SFRs) of 1--10 $M_\odot$ yr$^{-1}$, and a metal-poor interstellar medium (ISM) \citep{Nakajima2012ApJ...745...12N, Hagen2014ApJ...786...59H, Hagen2016ApJ...817...79H}. These distinctive properties make them ideal laboratories for investigating the early evolution of low-mass galaxies. Due to the strong resonant scattering and absorption of Ly$\alpha$ photons by neutral hydrogen (H\textsc{i}), LAEs provide valuable insights into the physical conditions of their surrounding environments and serve as effective tracers of ionized bubble structures during the epoch of reionization \citep{Hayes2310.3847/2041-8213/acee6a, Chen2024MNRAS.528.7052C, Witstok2024arXiv240816608W, Witstok2025MNRAS.536...27W}.

Morphological and size evolution serve as critical diagnostics for probing the physical processes underlying galaxy formation and evolution. For LAEs, early observations with the Hubble Space Telescope (HST) revealed that their rest-frame UV sizes are compact ($\sim$ 1 kpc) at 2 $<$ z $<$ 6 and show no statistically significant evolutionary trend across this redshift range \citep{Taniguchi2009ApJ...701..915T, Gronwall2011ApJ...743....9G, Malhotra2012ApJ...750L..36M, Jiang2013ApJ...773..153J, Hagen2014ApJ...786...59H, Hagen2016ApJ...817...79H, Kobayashi2016ApJ...819...25K, Paulino2018MNRAS.476.5479P}. 
In particular, \citet{Paulino2018MNRAS.476.5479P} quantified the size evolution as 
$R_{\rm e} \propto (1 + z)^{-0.21 \pm 0.22}$, consistent with negligible evolution. 
However, when accounting for the size-luminosity relation and selection biases, \citet{Shibuya2019ApJ...871..164S} derived a significantly steeper scaling, $R_{\rm e,circ} \propto (1 + z)^{-1.37\pm0.65}$, using a bias-controlled sample of LAEs over $z=2 \sim 7$. This stark contrast with earlier results highlights the sensitivity of inferred size evolution to sample selection and luminosity-dependent effects. The existing discrepancies indicate that the size evolution of LAEs remains poorly constrained, underscoring the need for further investigation with higher-resolution 
and higher-precision data--such as those enabled by JWST. Notably, a consistent picture emerges from studies of local analogs (e.g., Green Pea galaxies at z $\sim$ 0.3 \citealt{Yang2017ApJ...844..171Y, Kim2021ApJ...914....2K}) and LAEs at cosmic noon (1.7 < z < 3.3 \citealt{Kim2025arXiv250107548K}), which consistently reveal compact morphologies comparable to those of their high-redshift counterparts. This suggests that a compact size may be a persistent characteristic of LAEs across a wide range of redshifts.

The advent of the JWST has enabled a growing number of studies to investigate LAEs 
with unprecedented spatial resolution across a broad wavelength range. JWST/NIRCam provides extensive coverage 
up to $\sim$5 \textmu m, combining high sensitivity with sub-arcsecond resolution, 
thereby enabling detailed morphological analyses of LAEs at z $\gtrsim$ 3 \citep{Liu2024ApJ...966..210L, napolitano2024peering, Ning2024ApJ...963L..38N}. These investigations consistently reveal compact structures 
across multiple redshift regimes. For instance, \citet{Ning2024ApJ...963L..38N} measured 
a median circularized half-light radius of $R_{\rm e,circ} \sim 0.22^{+0.04}_{-0.04}$ kpc 
in the rest-frame UV for luminous LAEs ($L_{Ly\alpha} \sim 10^{42.4}$--$10^{43.4}$ erg~s$^{-1}$) at $z \sim 5.7$ in the COSMOS field, based on stacked NIRCam F115W and F150W images. 
Similarly, \citet{Liu2024ApJ...966..210L} reported a median $R_{\rm e,circ} \sim 0.36^{+0.04}_{-0.04}$ kpc in the rest-frame $V$ band for spectroscopically confirmed LAEs at $z \sim 3.1$ in the UDS field using F200W imaging. 
\citet{napolitano2024peering} further confirmed these results in the EGS field at 4.5 $<$ z $<$ 8.5, 
deriving a median size of $R_{\rm e} \sim 0.46^{+0.06}_{-0.06}$ kpc from F115W and F150W data 
in the rest-frame UV. Notably, \citet{Liu2024ApJ...966..210L} also derived a size--mass relation 
for their sample of 10 LAEs, finding good agreement with that of late-type galaxies at $z \sim 3$ \citep{van20143d}. 
Despite these valuable insights, current studies remain limited by small sample sizes and narrow redshift coverage. 
Such constraints hinder robust characterization of the full diversity and potential evolutionary trends 
in LAE morphologies. A comprehensive understanding of the size and structural evolution of LAEs 
therefore requires statistically robust, homogeneous samples spanning a wide redshift range. 

In this work, we present a uniformly selected sample of spectroscopically confirmed LAEs at 3 $\lesssim$ z $<$ 7 
across three deep extragalactic fields: GOODS-S, COSMOS, and UDS. Leveraging the latest JWST/NIRCam and MIRI imaging data, we conduct a systematic analysis of LAE size evolution in both the rest-frame UV and optical band. 
For the first time, we establish a robust size--stellar mass ($M_*$) relation for a large, homogeneous sample of high-z LAEs. We further examine the correlation between galaxy sizes measured in the UV and optical regimes. 
By utilizing JWST's unparalleled spatial resolution and incorporating multi-wavelength photometry, 
this study provides a statistically robust characterization of the size properties of LAEs 
across cosmic time.

This paper is structured as follows. In Section~\ref{sec:2}, we describe the selection of our LAE sample 
and the multi-wavelength JWST datasets used in this study. Section~\ref{sec:3} details the morphological analysis 
and spectral energy distribution (SED) modeling procedures. 
In Sections~\ref{sec:4} and~\ref{sec:5}, we present and discuss the results on size evolution, the size--mass relation, and the redshift evolution of the ratio of effective radii measured in different bands. 
A summary of our findings is provided in Section~\ref{sec:6}.
We adopt a concordance cosmology with $H_0 = 70 \, \text{km s}^{-1} \text{Mpc}^{-1}$, $\Omega_{\rm m} = 0.3$, and $\Omega_{\Lambda} = 0.7$. The initial mass function (IMF) assumed throughout this work is that of \citet{Chabrier2003PASP..115..763C}. Unless otherwise stated, all effective radii ($R_{\rm e}$) presented in this study 
correspond to the semi-major axes of the best-fit elliptical profiles 
derived from two-dimensional parametric surface brightness modeling.

\begin{figure*}[!t]
\centering
\includegraphics[width=1\textwidth]{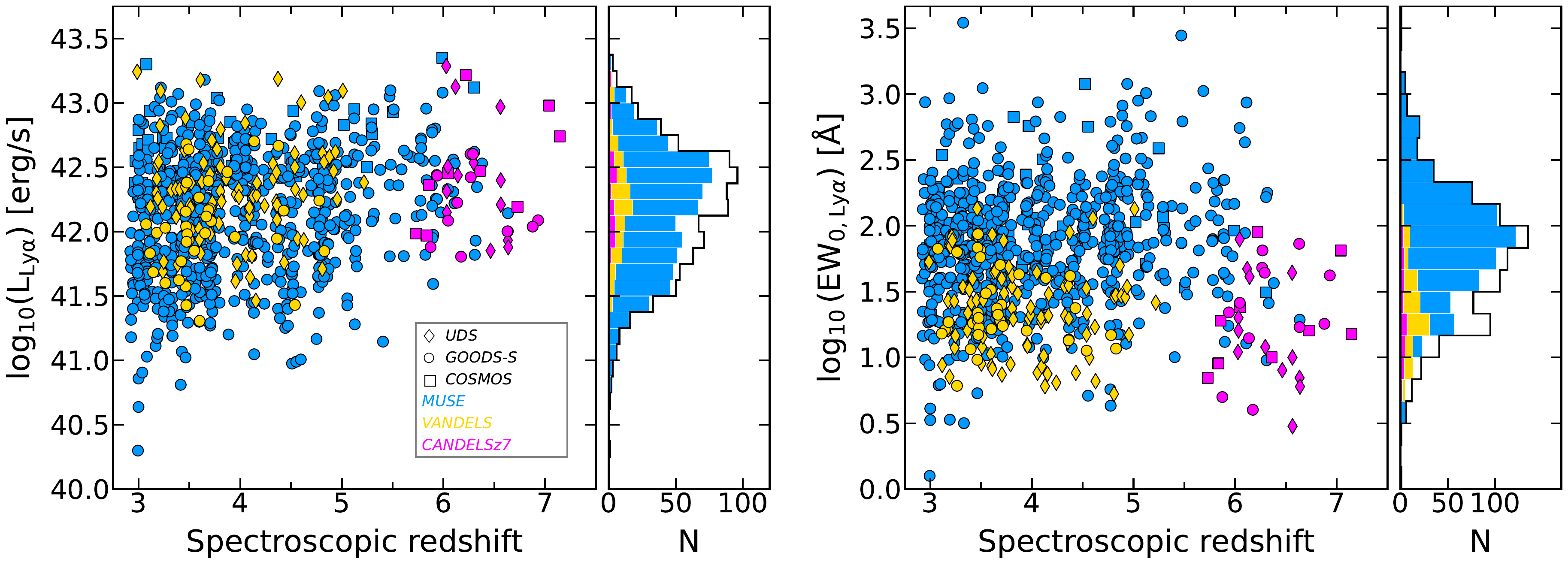} 
\caption{
Ly$\alpha$ luminosity (left) and rest-frame equivalent width (right) as a function of spectroscopic redshift for the 876 LAEs selected in this work, spanning the redshift range $z\sim3 \text{--} 7$. Symbol shapes indicate different fields (circle: GOODS-S, diamond: UDS, square: COSMOS), and colors correspond to different surveys (blue: MUSE, yellow: VANDELS, magenta: CANDELSz7).
}

\label{fig:Sample_image}
\end{figure*}

\section{DATA AND SAMPLE} \label{sec:2}

\subsection{JWST Data} \label{sec:2.1}

We utilize high-quality near-infrared imaging data from the JWST-SPRING project (Spatially Pixel-level Resolved Investigations into Nascent Galaxies with the James Webb Space Telescope), a public science initiative designed to provide the astronomical community with large-area, homogeneous, and deep imaging data from JWST’s NIRCam and MIRI instruments, precisely matched with existing HST imaging. In this work, we primarily use JWST/NIRCam imaging. The NIRCam data processing pipeline is summarized as follows:

Initial data processing was performed on raw, uncalibrated files using Stage 1 of the JWST calibration pipeline, which applies detector-level corrections through predefined parameter sets. Prior to Stage 2, we addressed 
"snowball" artifacts. In Stage 2, we followed the method described by \citet{schlawin2020jwst} to 
subtract "wisp" features and mitigate 1/f noise, and we applied dedicated masks to suppress artifacts evident in the mosaics, including persistence effects, "dragon's breath" , "ginko leaf" patterns, residual wisps, 
and other detector anomalies. For Stage 3, we implemented a specialized astrometric alignment 
and executed the \texttt{OutlierDetection} step, incorporating a customized outlier identification algorithm and localized background subtraction prior to resampling. The World Coordinate System (WCS) calibration was initially established using JWST/F150W imaging aligned to HST/F160W observations from the CANDELS survey \citep{Faber2011,2011ApJS..197...35G,2011ApJS..197...36K}, and this solution was then systematically propagated to all other NIRCam bands using the calibrated F150W data as the reference frame. For each observational dataset, individual mosaics were generated, and targeted background subtraction was applied to subregions of these mosaics to optimize the quality of the final data products.

In addition, we use MIRI data to validate stellar mass estimates derived from JWST/NIRCam and HST data alone. MIRI processing begins with Stage 1, which initializes data quality flags and applies standard detector corrections using default pipeline parameters. Stage 2 retains default settings but adds custom "super-background" subtraction and a stripe-removal algorithm. Stage 3 follows the NIRCam workflow, differing primarily in the astrometric reference catalog: for MIRI, we built a catalog from F444W mosaics, supplemented with HST F160W sources in areas lacking F444W coverage, and performed two TweakReg passes to align all exposures to an absolute WCS. 
For further details on the project and data products, see the JWST-SPRING website\footnote{\url{http://groups.bao.ac.cn/jwst_spring/}} and the overview paper by Liu et al. (in preparation).

\subsection{The Parent Sample} \label{sec:2.2}

We collected a large sample of spectroscopically confirmed LAEs in the GOODS-S, UDS and COSMOS fields over the 
redshift range of z $\sim$ 3 \text{--} 7, based on data from the MUSE \citep{Herenz2017, Bacon2017}, VANDELS \citep{McLure18} and CANDELSz7 \citep{Pentericci18} spectroscopic surveys. 

\subsubsection{MUSE Survey}\label{sec:2.2.1}

The MUSE survey, conducted using the VLT/MUSE instrument \citep{2014Msngr.157...13B}, consists of 
two components: MUSE-Wide \citep{Herenz2017, 2019A&A...624A.141U} and MUSE-Deep \citep{Bacon2017, 2017A&A...608A...2I, Bacon23}. MUSE-Wide targets LAEs at redshifts 2.9 $<$ z $<$ 6.7 in the Chandra Deep Field South (CDFS) 
and COSMOS fields. Ly$\alpha$ line fluxes are measured using asymmetric Gaussian fitting, and rest-frame UV continuum levels are derived from HST broad-band data to compute equivalent widths (EWs).
LAEs are selected from the \citet{Kerutt22} catalog based on a \texttt{Confidence level from QtClassify} = 2 or 3, 
corresponding to an error rate of $\lesssim 10$\%. 
In contrast, MUSE-Deep focuses on the Hubble Ultra Deep Field (HUDF) with ultra-deep exposures 
ranging from 10 to 141 hours. There, Ly$\alpha$ fluxes and EWs are measured 
using \texttt{pyPlatefit} \citep{Bacon23}, employing skewed or double-peaked Gaussian models; 
these measurements are validated against narrowband (NB) imaging. 
LAEs are retained from the Data Release 2 (DR2) catalog \citep{Bacon23} with a \texttt{ZCONF} = 2 or 3 -- corresponding to a signal-to-noise ratio (S/N) of at least 5 or 7 -- based on a combination of automated redshift determination (\texttt{ORIGIN} \citep{Mary2020A&A...635A.194M}, \texttt{ODHIN} \citep{bacher2017methodes,Bacon23}) and expert visual inspection.
We selected galaxies with secure Lyman-alpha emission-line detections ($S/N >3$)
based on data from \citet{Kerutt22} and \citet{Bacon23}.
For sources common to both the MUSE-deep and MUSE-wide surveys, we prioritized
the deeper MUSE-deep observations \citep{Bacon23} to ensure higher spectral sensitivity and data consistency.

\subsubsection{VANDELS Survey}\label{sec:2.2.2}

The VANDELS survey, conducted with VLT/VIMOS \citep{LeFevre2003}, provides deep optical spectroscopy (4800\text{--}9800 \AA) targeting high-redshift galaxies in the Chandra Deep Field South (CDFS) and the UKIDSS Ultra Deep Survey (UDS) 
fields. Our sample of LAEs is drawn from the VANDELS DR4 catalog \citep{Garilli2021A&A...647A.150G}, 
in which spectroscopic redshifts and line parameters were measured using the \texttt{EZ} software within the \texttt{PANDORA} environment \citep{garilli2010ez}. Ly$\alpha$ equivalent widths were calculated following the method of 
\citet{2010ApJ...711..693K}: continuum regions are defined blueward (1120\text{--}1180 \AA) and redward (1225\text{--}1255 \AA) of the Ly$\alpha$ line; the flux between these boundaries is integrated; and the red-side continuum is used as the baseline. We selected LAEs with a \texttt{Redshift confidence flag} = 3 or 4 and a 
\texttt{Ly$\alpha$ goodness-of-fit flag} = 1 from the Gaussian-fit catalogs of \cite{Talia23}. 
We require \texttt{Ly$\alpha$ Line flux} $>$ 0 to ensure emission-line nature, as negative flux values 
indicate Ly$\alpha$ in absorption.

\subsubsection{CANDELSz7 Survey}\label{sec:2.2.3}

The CANDELSz7 survey \citep{Pentericci18} identifies LAEs at redshifts 6 $<$ z $<$ 7 
within the CANDELS fields \citep{2011ApJS..197...35G,2011ApJS..197...36K}, 
using observations from VLT/FORS2 equipped with the 600Z grism (wavelength coverage: 8000\text{--}10000 \AA, spectral resolution R $\sim$ 1390). Ly$\alpha$ line fluxes are measured by direct integration of the 1D spectra, 
under the assumption of negligible slit losses for compact sources. Equivalent widths are computed 
using HST photometry to estimate the rest-frame UV continuum level -- specifically the UV spectral slope ($\beta$) -- 
at 1300 $\times$ (1 + z) \AA. For sources undetected in emission, 3$\sigma$ upper limits on EW are derived from sensitivity simulations performed at  z= 6.0 (i\text{--}dropouts) and z = 6.9 (z\text{--}dropouts). 
LAEs are selected from the the \citet{Pentericci18} catalog with Ly$\alpha$ \texttt{flux} $>$ 0 and \texttt{comm.} = Ly$\alpha$, ensuring robust identification of Ly$\alpha$ emission. 

\begin{deluxetable*}{cccc}
\tablecaption{The Final LAE Sample}
\tablewidth{0pt}
\setlength{\tabcolsep}{7mm}
\label{result}
\tablehead{
LAE Sample     & $z$       & Number of LAEs & Reference Catalog
}
\startdata
MUSE-Deep GOODS-S & $3.0 \sim 6.7$ & 232 & \citet{Bacon23} \\
MUSE-Wide GOODS-S & $3.0 \sim 6.5$ & 434 & \citet{Kerutt22} \\
MUSE-Wide COSMOS      & $3.0 \sim 6.5$ & 54  & \citet{Kerutt22} \\
VANDELS GOODS-S   & $3.0 \sim 4.8$ & 37  & \citet{Talia23} \\
VANDELS UDS           & $3.0 \sim 5.2$ & 85  & \citet{Talia23} \\
CANDELSz7 GOODS-S & $6.0 \sim 7.0$ & 12  & \citet{Pentericci18} \\
CANDELSz7 COSMOS      & $6.0 \sim 7.0$ & 9  & \citet{Pentericci18} \\
CANDELSz7 UDS         & $6.0 \sim 7.0$ & 13  & \citet{Pentericci18} \\  \hline
Total                 & $3.0 \sim 7.0$         & 876 & \dots \\
\hline
\enddata
\tablecomments{
The redshift ranges represent the final selected spectroscopic redshifts for the LAEs from the corresponding reference catalogs. The number of LAEs reported is the result of applying the filtering conditions described in Section \ref{sec:2.1}, followed by de-duplication and cross-matching with our JWST image data.
}
\end{deluxetable*}

\subsection{Final Sample of LAEs} \label{sec:2.3}

From the parent sample, we initially selected 1260 LAE candidates. After careful visual inspection to remove sources lacking JWST coverage -- insufficient for SED fitting and 
missing coverage at the corresponding rest-frame wavelengths required for size measurements -- 
as well as heavily contaminated objects, the final sample comprises 876 LAEs. This sample is summarized 
in Table~\ref{result}, and Figure~\ref{fig:Sample_image} presents their Ly$\alpha$ 
luminosity and rest-frame equivalent width as a function of spectroscopic redshift.

Additionally, we addressed potential contamination from active galactic nuclei (AGN) in our final LAE sample. While the other spectroscopic surveys (excluding MUSE-Deep) had already removed AGN or provided AGN classification flags, we applied the AGN exclusion criteria from \citet{bacon2021A&A...647A.107B} to LAEs selected from \citet{Bacon23}. Specifically, we excluded potential AGN by cross-matching the MUSE-Deep spectroscopic sample with the AGN catalog compiled by \citet{luo2017ApJS..228....2L}.

\begin{figure}[t!]
\centering
\includegraphics[width=1\columnwidth]{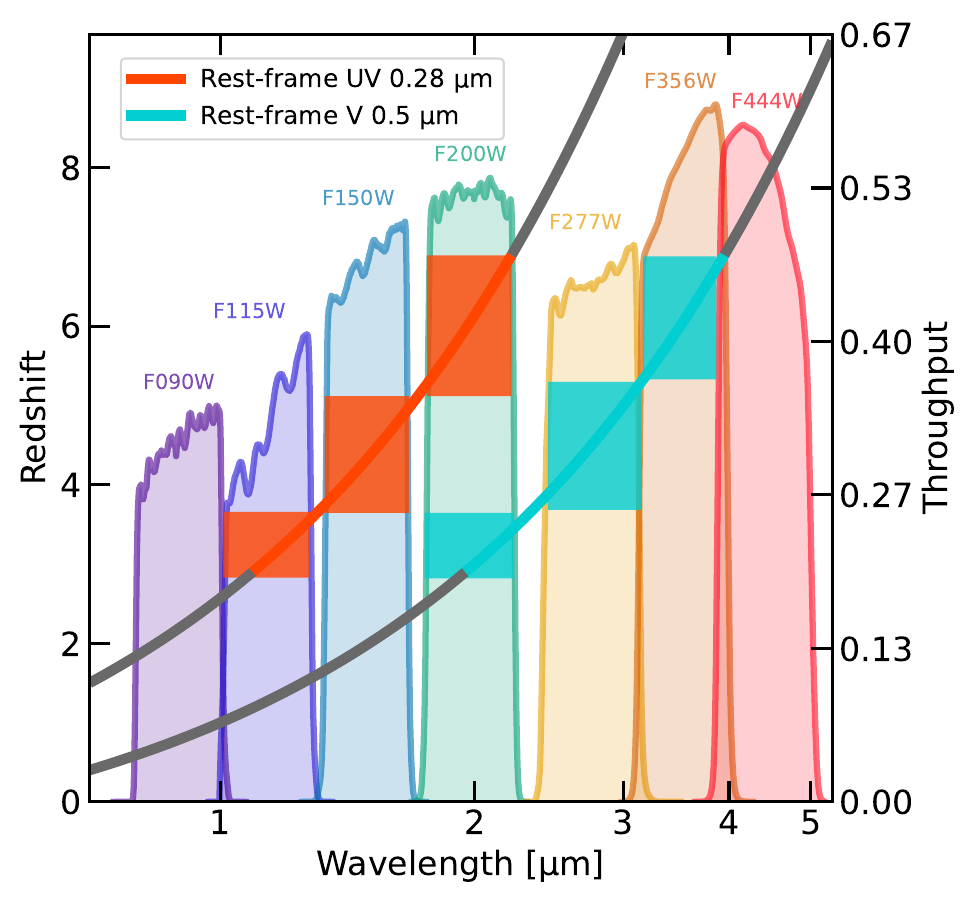} 
\caption{
JWST/NIRCam broad-band filters used for measuring sizes in the rest-frame UV-band (0.28 \textmu m, orange) and rest-frame V-band (0.5 \textmu m, cyan).
}
\label{fig:filter_image}
\end{figure}

\section{METHODS} \label{sec:3}

\subsection{Size Measurement} \label{sec:3.1}

Our objective is to determine the sizes of these LAEs in the rest-frame UV and V bands separately. To accomplish this, for each LAE we selected the JWST/NIRCam filter whose rest-frame central wavelength 
is closest to 0.28 \textmu m for the UV band and to 0.5 \textmu m for the V band, based on its spectroscopic redshift. Figure \ref{fig:filter_image} illustrates how the selected NIRCam filters map to the rest-frame UV and V bands 
across different redshift intervals. 

We utilized four separate S\'{e}rsic fitting software packages -- \texttt{Pysersic} \citep{pasha2023pysersic}, \texttt{PetroFit} \citep{geda2022petrofit}, \texttt{Statmorph} \citep{rodriguez2019optical}, and \texttt{GalfitM} \citep{haussler2013megamorph, vika2013megamorph} -- to perform independent S\'{e}rsic fitting. 
Each software package has distinct strengths. 
\texttt{Pysersic} employs Bayesian methods to model galaxy morphologies, leveraging advanced computational techniques to rapidly estimate uncertainties in fitted parameters. 
\texttt{PetroFit} computes Petrosian quantities and performs S\'{e}rsic fitting, and is designed for high-precision photometry and segmentation, making it well-suited for diverse galaxy morphologies. 
\texttt{Statmorph} calculates non-parametric morphological diagnostics (e.g., Gini, $M_{20}$), and also 
includes S\'{e}rsic fitting and other structural analysis tools. 
\texttt{GalfitM}, an upgraded version of \texttt{GALFIT} \citep{peng2002detailed, peng2010detailed}, is primarily designed for simultaneous multi-band fitting. However, for single-band fitting, it yields statistically consistent results with \texttt{GALFIT}, achieving comparable accuracy and reliability within measurement uncertainties \citep{haussler2013megamorph}. We use \texttt{GalfitM} exclusively for single-band image fitting in this work. 
All four software packages employ S\'{e}rsic profile fitting \citep{sersic1963influence, sersic1968atlas} to measure galaxy sizes using the effective radius $R_{\rm e}$ as the semi-major axis enclosing half the total light.
Point-spread function (PSF) deconvolution is applied to all size measurements using PSFs from the JWST-SPRING library, which are reconstructed from stacked, unsaturated bright stars in each field.

\begin{figure*}[t!]
\centering
\includegraphics[width=1\textwidth]{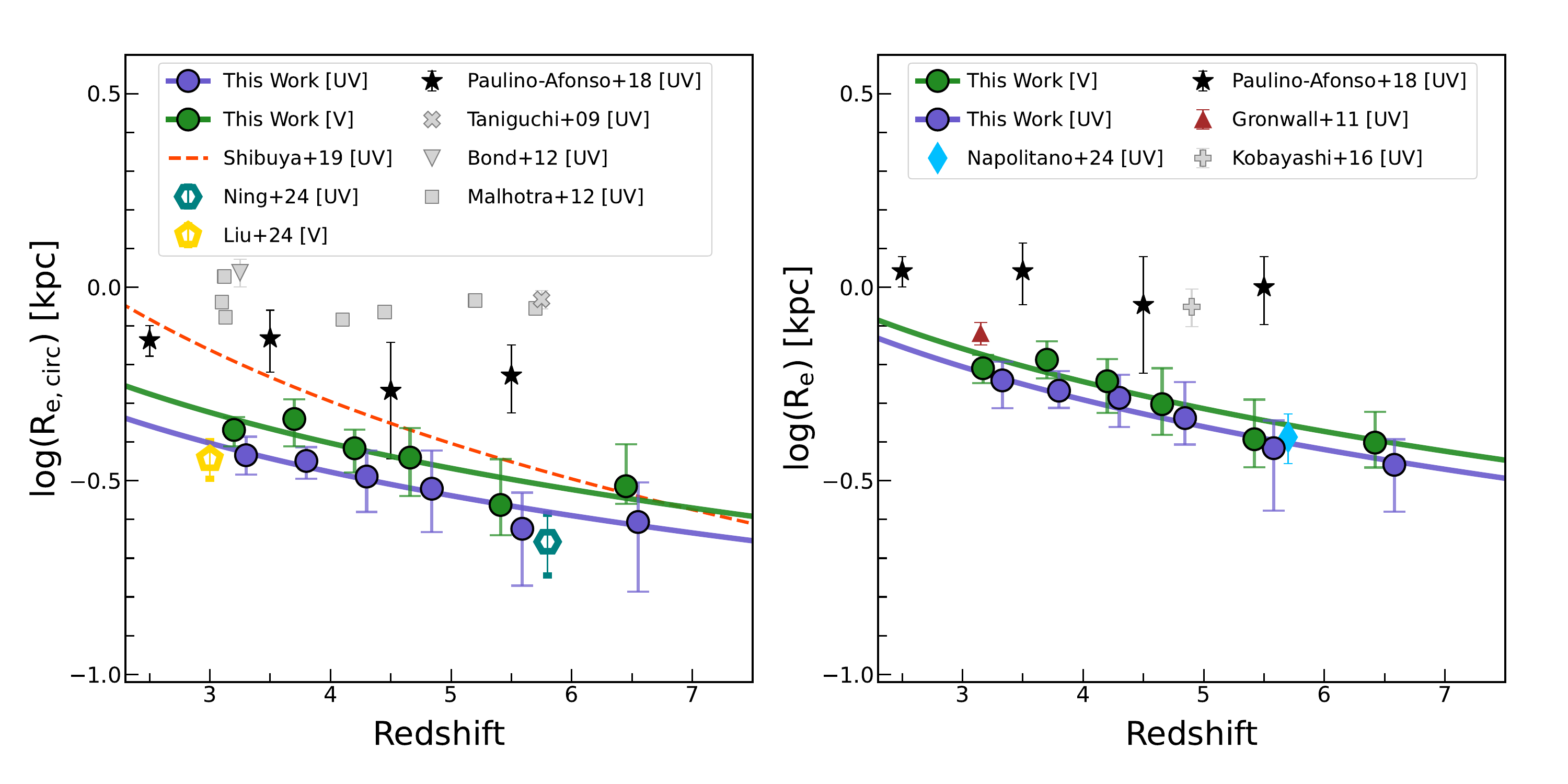} 
\caption{
Effective radii as a function of redshift for LAEs in the rest-frame UV and V bands over the redshift range 
$\sim3<z<7$. 
\textbf{Left panel:} Effective radii are circularized as $R_{\rm e,circ} = R_{\rm e} \sqrt{b/a}$, where $b/a$ is the axis ratio. Purple and green circles represent the median $R_{\rm e,circ}$ values in the UV and V bands, 
respectively, within redshift bins. 
Solid lines show the best-fit $R_{\rm e,circ}$\text{--}$z$ relations for each band. 
Vertical error bars indicate the uncertainty in the median for each redshift bin. 
The red dashed line shows the UV-band size evolution ($R_{\rm e, circ} \propto (1 + z)^{-1.37 \pm 0.65}$) 
for LAEs at $2 < z < 7$ with $L_{UV} = (0.12-1)L_{\rm z=3}^{*}$, measured from rest-frame wavelengths 
of 0.15--0.3 \textmu m by \citet{Shibuya2019ApJ...871..164S}, based on data 
from $V_{606}, I_{814}, J_{125}, H_{160}$, or the co-added WFC3 bands. 
Black filled stars denote LAEs in the COSMOS field ($i_{AB} < 25$, 2 $<$ z $<$ 6) 
analyzed using HST/ACS F814W imaging by \citet{Paulino2018MNRAS.476.5479P}. 
The yellow open pentagon marks spectroscopically confirmed LAEs in the UDS field at $z = 3.1$, 
characterized using JWST/NIRCam F200W imaging by \citet{Liu2024ApJ...966..210L}. 
The teal open hexagon represents high-luminosity LAEs ($L_{\mathrm{Ly\alpha}} \sim 10^{42.4} \,\text{--}\, 10^{43.4}$ erg~s$^{-1}$) in the COSMOS field at $z \sim 6$, measured from stacked JWST/NIRCam F115W and F150W images by 
\citet{Ning2024ApJ...963L..38N}. 
Gray symbols -- squares \citep{Malhotra2012ApJ...750L..36M}, 
inverted triangle \citep{Bond2012ApJ...753...95B}, 
cross \citep{Kobayashi2016ApJ...819...25K} (right panel), and x-mark \citep{Taniguchi2009ApJ...701..915T} 
-- represent HST-based measurements derived from direct photometric parameters, 
rather than from S\'{e}rsic profile fitting.
\textbf{Right panel:} Purple, green, and black points, along with the corresponding best-fit solid lines, are 
taken from the same dataset as in the left panel, whereas the sizes refer to effective semi-major axes. 
For comparison, the brown filled triangle represents LAEs in the ECDF\text{--}S at $z = 3.1$, 
measured from HST/ACS V-band imaging by \citet{Gronwall2011ApJ...743....9G}. 
The cyan filled diamond shows the median size of LAEs in the EGS field at 4 $<$ z $<$ 7, 
selected based on detected Ly$\alpha$ emission with S/N $ > $ 3, and derived from JWST/NIRCam F115W and 
F150W imaging by \citet{napolitano2024peering}. 
All data points are horizontally offset for clarity.
}
\label{fig:size_image}
\end{figure*}

By systematically comparing $R_{\rm e}$ measurements from four independent software packages applied 
to consistently masked images with contaminant removal, we confirmed a high degree of convergence 
in galaxy size estimates across different methodologies. A statistical evaluation of the fitting methods 
reveals that \texttt{PetroFit} exhibits fewer outliers compared to the other tools. 
Following the strategy outlined in \citet{Ren25}, we adopt the \texttt{PetroFit} result 
if it falls within the range defined by the other three measurements. 
If \texttt{PetroFit} fails to return a valid fit or its result lies outside this range, 
we instead adopt the median value from the remaining three. 
This approach effectively reduces the impact of outliers that can arise from relying on a 
single fitting software. 

\citet{Ren2025arXiv250723654R} reported that over 40$\%$ of the LAEs in the GOODS-S sample 
are late-stage merging systems. To ensure robust size measurements for these objects, we performed multi-component decomposition using the \texttt{FitMulti} algorithm in the \texttt{Pysersic} package. 
For LAEs with clearly resolved merging components, we selected the component spatially closest to the Lyman-alpha emission as the representative LAE. Representative fitting diagnostics, including residual comparisons and consistency of fitted parameters across different software packages, are presented in Appendix \ref{size_fitting}. 
After excluding sources with incomplete band coverage or poor model fits, the final sample for analysis 
includes 844 ($\sim$96.3\%) sources with measured $R_{\rm e}$ in the UV band and 839 ($\sim$95.4\%) sources in the V band, with 807 $\sim$(92.1\%) sources having $R_{\rm e}$ measurements in both bands.

\subsection{Photometry and SED Fitting} \label{sec:3.2}

Leveraging the rich multi-wavelength data from the JWST-SPRING project combining JWST and HST observations, we perform multi-band photometry and SED modeling for these LAEs. In this work, since the JWST/NIRCam bands cover the HST WFC3/IR passbands and offer superior spatial resolution and depth, we use only HST/ACS (F606W, F814W) and JWST/NIRCam (F090W, F115W, F150W, F200W, F277W, F356W, F444W) data. Additionally, if a source is detected in JWST/MIRI F770W, we include this mid-infrared band in the analysis. Total fluxes for the LAEs are measured using the Python package \texttt{photutils} \citep{larry_bradley_2024_13989456}, based on PSF-matched images with contaminating sources masked. SED modeling is performed with the \texttt{CIGALE} \citep{Boquien2019A&A...622A.103B, Yang2020MNRAS.491..740Y, Yang2022ApJ...927..192Y} fitting code, with redshifts fixed to spectroscopically confirmed values. We adopt the \texttt{sfhdelayed} star formation history model, allowing the age of the late burst population to vary from 1 to 50 Myr and the age of the main stellar population from 10 to 3000 Myr. Stellar population synthesis is carried out using the \texttt{BC03} module \citep{Bruzual2003MNRAS.344.1000B} with a Chabrier IMF \citep{Chabrier2003PASP..115..763C}. Dust attenuation is modeled using the \texttt{dust-modified-starburst} law \citep{Calzetti2000ApJ...533..682C} with $E(B - V)_{\mathrm{lines}}$ ranging from 0 to 0.6. To account for differential dust extinction between young and old
 stellar populations, we adopted a scaling factor of 0.44 for the stellar continuum $E(B - V)$ 
\citep{Calzetti2000ApJ...533..682C, Charlot2000ApJ...539..718C, Wild2011MNRAS.417.1760W}.

Recently, \citet{Iani2024ApJ...963...97I} demonstrated that JWST/MIRI F560W photometry has only a minimal impact on stellar mass estimates derived from SED fitting for LAEs using JWST/NIRCam and MIRI data. In this study, we specifically investigated the impact of longer-wavelength JWST/MIRI F770W data on stellar mass determination by performing comparative SED fitting for LAEs in the COSMOS and UDS fields with and without F770W photometry. As detailed in Appendix \ref{stellar_mass}, the results show that excluding F770W data 
leads to a negligible difference of stellar mass ($<$ 0.1 dex), indicating that the absence of JWST/MIRI observations introduces minimal systematic bias in the stellar mass measurements of LAEs. Therefore, stellar mass estimates for galaxies in our sample lacking F770W coverage remain highly reliable.

We will provide readers with a machine-readable table containing the full sample of 876 LAEs and their basic physical properties. A representative example of this table (Table 3) is presented in Appendix~\ref{LAE_catalog}. These sources were selected sequentially from different observational fields to ensure representative coverage across the various surveys.

\section{RESULTS} \label{sec:4}

\subsection{Size Evolution} \label{sec:4.1}

\begin{nolinenumbers}
\begin{table}[t!]
\caption{Number of LAEs in different redshift bins used for the size evolution analysis.}
\label{Table2}
\centering
\setlength{\tabcolsep}{1.5mm}
\renewcommand{\arraystretch}{1.2}
\small
\begin{tabular}{>{\centering}m{1.9cm}
                >{\centering\arraybackslash}m{1.2cm}
                >{\centering\arraybackslash}m{1.2cm}
                >{\centering\arraybackslash}m{1.2cm}
                >{\centering\arraybackslash}m{1.2cm}} 
\hline\hline
\multicolumn{1}{c}{Range} & 
\multicolumn{1}{c}{$R_{\rm e, UV}$} & 
\multicolumn{1}{c}{$N_{UV}$} &
\multicolumn{1}{c}{$R_{\rm e, V}$} & 
\multicolumn{1}{c}{$N_{V}$} \\ 
\multicolumn{1}{c}{} &  
\multicolumn{1}{c}{(kpc)} & 
\multicolumn{1}{c}{} &
\multicolumn{1}{c}{(kpc)} & 
\multicolumn{1}{c}{} \\ 
\hline
$3.0 < z < 3.5$ & $0.57^{+0.59}_{-0.37}$ & 273(97\%) & $0.62^{+0.58}_{-0.36}$ & 270(96\%) \\ 
\hline 
$3.5 < z < 4.0$ & $0.54^{+0.54}_{-0.33}$ & 200(97\%) & $0.65^{+0.59}_{-0.35}$ & 198(96\%) \\ 
\hline 
$4.0 < z < 4.5$ & $0.52^{+0.36}_{-0.29}$ & 125(97\%) & $0.57^{+0.48}_{-0.30}$ & 123(95\%) \\ 
\hline 
$4.5 < z < 5.0$ & $0.46^{+0.48}_{-0.26}$ & 125(95\%) & $0.50^{+0.48}_{-0.27}$ & 130(99\%) \\ 
\hline 
$5.0 < z < 6.0$ & $0.38^{+0.34}_{-0.21}$ & 80(92\%) & $0.41^{+0.46}_{-0.21}$ & 79(91\%) \\ 
\hline 
$6.0 < z < 7.0$ & $0.35^{+0.25}_{-0.16}$ & 41(100\%) & $0.40^{+0.24}_{-0.15}$ & 39(95\%) \\ 
\hline
\end{tabular}
\end{table}
\end{nolinenumbers}

\begin{figure*}[t!]
\centering
\includegraphics[width=1\textwidth]{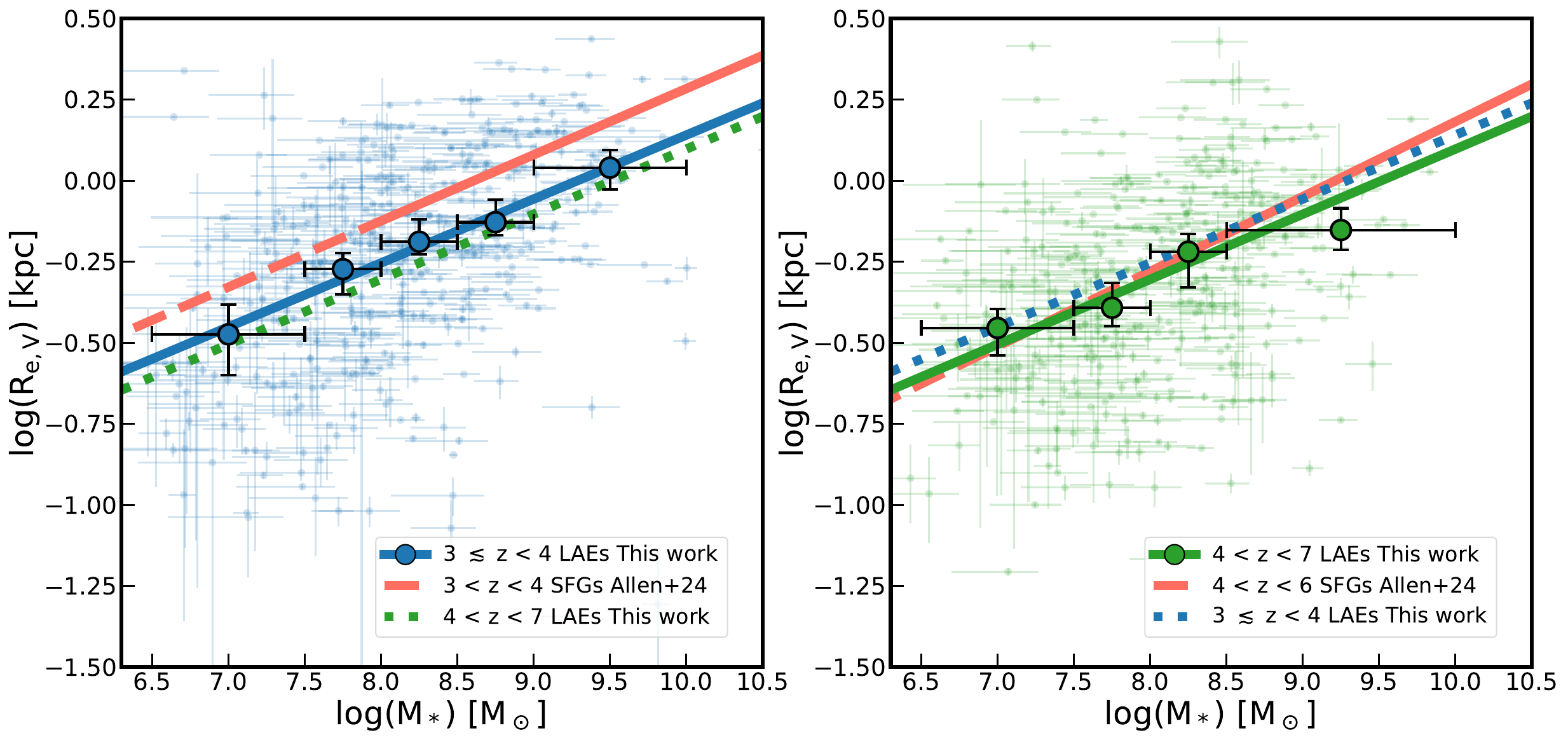} 
\caption{
Rest-frame V-band size-mass relations for LAEs at redshifts $3 \lesssim z < 4$ (left panel) 
and $4 < z < 7$ (right panel). \textbf{Left panel:} The solid blue line shows the best-fit relation for our LAE sample. Blue filled circles represent the median sizes in each stellar mass bins, with vertical error bars indicating the uncertainties and horizontal bars showing the bin widths. 
\textbf{Right panel:} The solid green line shows the best-fit relation for LAEs at higher redshift. Green filled circles with error bars have the same meaning as in the left panel. 
Orange lines in both panels show the size-mass relations for star-forming galaxies (SFGs) from 
\citet{Allen2024arXiv241016354A} at $3 < z < 4$ (left) and $4 < z < 6$ (right).  Solid orange segments indicate the fitted range of the literature data, while dashed extensions represent extrapolations beyond that range.
}
\label{fig:size_mass}
\end{figure*}

Figure \ref{fig:size_image} presents the evolution of the rest-frame UV effective radius ($R_{\rm e,UV}$, at $\sim$ 0.28 \textmu m) and the rest-frame V-band effective radius with redshift for our LAEs sample. 
We adopted two distinct size characterization approaches, which not only facilitate direct comparison with prior studies but also provide a comprehensive and unbiased description of the size properties of LAEs. 
In Figure \ref{fig:size_image}, the left panel shows the redshift evolution of the circularized LAEs radii calculated by multiplying by the square root of the axis ratio ($R_{\rm e,circ} \equiv R_{\rm e} \times \sqrt{q}$), while the right panel shows the evolution of the elliptical semi-major axis. For our LAEs sample, axial ratios typically lie in the range of 0.45--0.50, which is broadly consistent with 
the axial ratio distribution reported by \citet{Paulino2018MNRAS.476.5479P}. 
Several representative examples are provided in Appendix \ref{axial ratio}.

Our analysis reveals that LAEs exhibit minimal evolution in their $R_{\rm e}$ in 
both rest-frame UV and bands over the redshift range of 3 $\lesssim$ z $<$ 7. 
Specifically, in the rest-frame UV, the median size increases from $R_{\rm e} = 0.32_{-0.09}^{+0.11}$\,kpc 
at $z \sim 7$ to $R_{\rm e} = 0.62_{-0.15}^{+0.18}$\,kpc at $z \sim 3$ with decreasing redshift. 
In the rest-frame V band, the size increases from $R_{\rm e} = 0.36_{-0.16}^{+0.23}$\,kpc 
at $z \sim 7$ to $R_{\rm e} = 0.71_{-0.30}^{+0.47}$\,kpc at $z \sim 3$. 
Table \ref{Table2} summarizes the rest-frame UV and V-band $R_{e}$ measurements for our LAE sample, 
presenting median values with 16th--84th percentile ranges, along with the number of sources 
and their percentage representation relative to the parent sample in each redshift bin. 

The size evolution of galaixes with redshift is commonly described by a simple power-law relation. 
Following previous studies \citep[e.g.,][]{Paulino2018MNRAS.476.5479P, Shibuya2019ApJ...871..164S}, 
we model this evolution as: 
\begin{equation}
    \frac{R_{\rm e}}{\rm kpc} = \alpha \cdot (1 + z)^{\beta}
\end{equation}
where $\alpha$ and $\beta$ are free parameters. 
The fitting is based on the median values in each redshift bin, as listed in Table \ref{Table2}. 
We find that $R_{\rm e,UV} = (2.21\pm0.36) \times (1 + z)^{-0.91\pm0.10}$ and $R_{\rm e,V} = (2.54\pm0.78) \times (1 + z)^{-0.93\pm0.18}$. These parametric relations reveal mild evolutionary trends in the rest-frame UV and 
optical sizes of LAEs over the redshift range $3 \lesssim z < 7$. 

Additionally, Figure \ref{fig:size_image} compiles the size measurements of LAEs from 
the literature \citep{Taniguchi2009ApJ...701..915T, Gronwall2011ApJ...743....9G, Bond2012ApJ...753...95B, Malhotra2012ApJ...750L..36M, Kobayashi2016ApJ...819...25K, Paulino2018MNRAS.476.5479P, Shibuya2019ApJ...871..164S, Liu2024ApJ...966..210L, Ning2024ApJ...963L..38N, napolitano2024peering}, including previous studies based on the HST and recent observations from the JWST. It can be seen that within the redshift range 3 $\lesssim$ z $<$ 7, the size evolution of LAEs in our sample is generally consistent with previous observations, although there are some minor differences. 
These differences may be attributed to variations in sample selection or data quality and will be discussed 
in Section \ref{sec:5.1}.

\subsection{Size-Mass Relation} \label{sec:4.2}

Figure~\ref{fig:size_mass} presents the relationships between rest-frame V-band sizes 
and stellar masses for LAEs. 
To investigate potential redshift-dependent trends and enable direct comparisons 
with previous studies, we divide the sample into two redshift bins: a low-redshift 
subset ($3 \lesssim z < 4$; left panel) 
and a high-redshift subset ($4 < z < 7$; right panel). The size-mass relation 
in each redshift bin is fitted using the form:
\begin{equation}
    \rm log \frac{R_{\rm e}}{\rm kpc} = \alpha \cdot \rm log \frac{M_{*}}{M_{\odot}} + \rm log A
\end{equation}
where $\alpha$ and $A$ are free parameters. 

Our results show that the slopes of the size-mass relation for LAEs are 
consistent across two redshift intervals, with $\alpha \sim 0.2$ for both 
$3 \lesssim z < 4$ and $4 < z < 7$. In the lower-redshift bin ($3 \lesssim z < 4$), 
rest-frame V-band sizes increase from $\sim 0.37$ kpc at $\log(M_{*}/M_\odot) = 7$ to $\sim 1.18$ kpc 
at $\log(M_{*}/M_\odot) = 9.5$. At higher redshifts ($4 < z < 7$), the sizes 
grow from $\sim 0.31$ kpc to $\sim 0.98$ kpc over the same mass range. 
The best-fit relations are $\log R_{\rm e,V} = (0.20 \pm 0.01) \times (\log M_{*}/M_\odot) - (1.83 \pm 0.12)$ 
for the low-redshift bin, and $\log R_{\rm e,V} = (0.20 \pm 0.03) \times (\log M_{*}/M_\odot) - (1.91 \pm 0.23)$ for the high-redshift bin, respectively. 

Comparing our results with recent JWST-based studies of the size-mass relation for 
low-mass SFGs by \citet{Allen2024arXiv241016354A}, 
we find that the evolutionary trend of LAEs is broadly consistent with that of low-mass SFGs, albeit with minor differences. These discrepancies are discussed in detail in Section \ref{sec:5.2}.

\subsection{Comparison Between Rest-Frame Optical and UV Sizes} \label{sec:4.3}

\begin{figure}[t!]
\centering
\includegraphics[width=1\columnwidth]{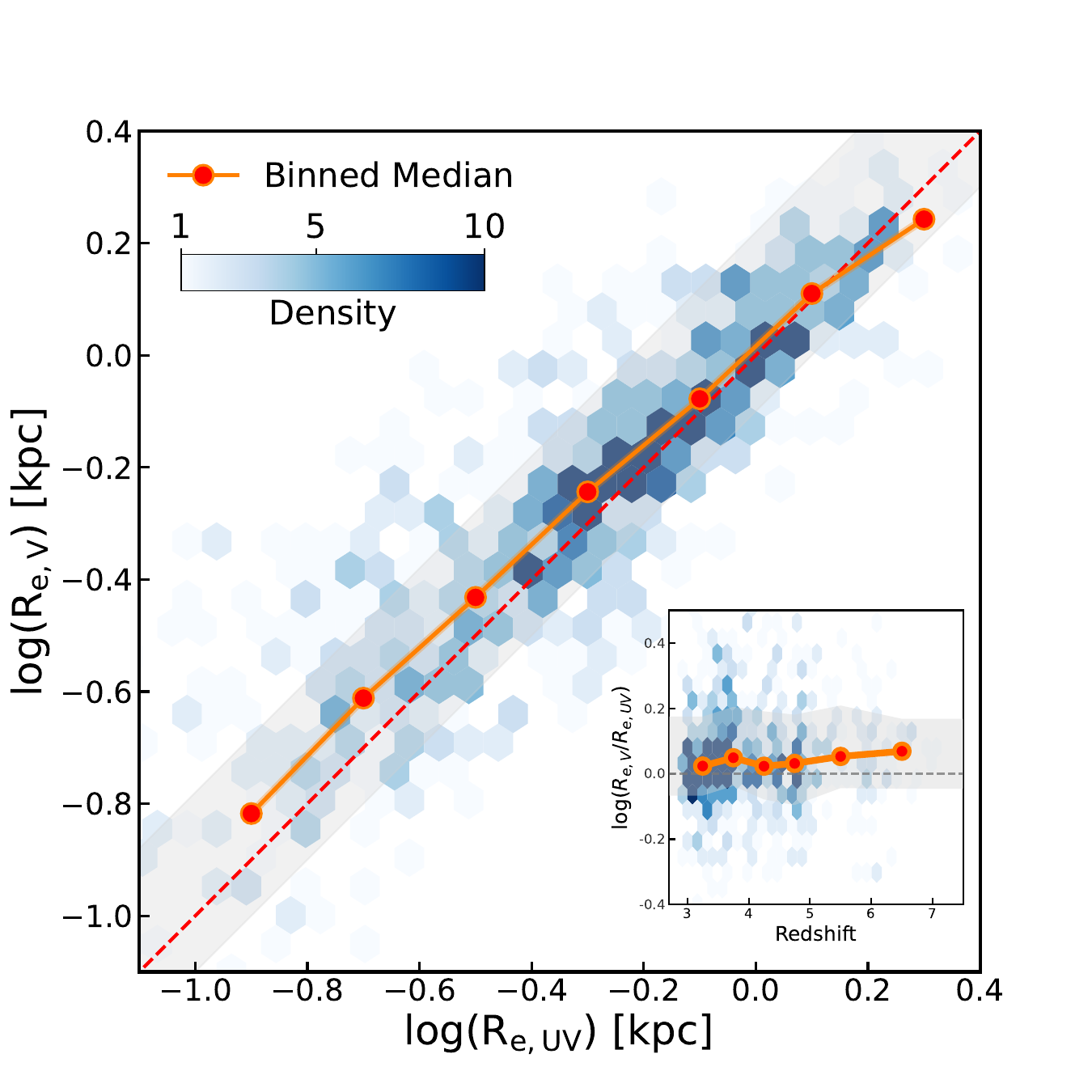} 
\caption{
Comparison between the effective radii of LAEs measured in the rest-frame UV and optical bands, 
for sources with good S\'{e}rsic profile fits in both bands. 
Each hexagonal bin is color-coded according to the number of sources it contains, 
as indicated by the colorbar. The red dashed line represents the one-to-one relation. 
The inset panel in the bottom-right corner shows the median logarithmic ratio of 
optical to UV effective radii ($\log(R_{\rm e,V}/R_{\rm e,UV})$). 
}
\label{fig:size_compare}
\end{figure}

As shown in Section \ref{sec:4.1}, the rest-frame UV and optical sizes of LAEs exhibit similar evolutionary trends and are comparable in scale over the redshift range $3 \lesssim z < 7$.
Figure \ref{fig:size_compare} compares the rest-frame UV and optical effective radii
for LAEs with successful S\'{e}rsic profile fits in both bands.
Most data points lie close to the one-to-one line, indicating general consistency
between the UV and optical sizes. This agreement is further supported by the small median logarithmic offset, 
$\log(R_{\rm e,V}/R_{\rm e,UV}) \approx 0.03$, suggesting negligible systematic size differences between the two bands. 
This constancy implies that LAEs have weak or negligible color gradients across the rest-frame UV to optical wavelengths over the studied redshift range.

\section{DISCUSSIONS} \label{sec:5}

\subsection{Comparison with Previous Size Measurements} \label{sec:5.1}

As shown in Figure \ref{fig:size_image}, the effective (half-light) radii of our 
LAEs - both along the semi-major axis and after circularization - are in close agreement 
with the latest JWST-based measurements reported in recent studies ($R_{\rm e}$: \citet{napolitano2024peering}; $R_{\rm e,circ}$: \citet{Liu2024ApJ...966..210L, Ning2024ApJ...963L..38N}). 
Notably, the measurements from \citet{Ning2024ApJ...963L..38N} and \citet{Liu2024ApJ...966..210L} 
at redshifts $z=5.7$ and $z=3.1$ are approximately $\sim 0.1$ dex smaller than ours, 
a discrepancy that may stem from their relatively small sample sizes.

When comparing our results with those based on HST data, we find that the measurements by \citet{Gronwall2011ApJ...743....9G} are generally consistent with ours, though they are slightly larger by $\sim$0.08 dex. This small offset may stem from the lower angular resolution and sensitivity of HST compared to JWST, potentially leading to the omission of smaller galaxies or an overestimation of their sizes. Such effects are likely more significant for LAEs, given their compact morphologies and greater sensitivity to instrumental limitations. 
Furthermore, we find that the size measurements reported by \citet{Paulino2018MNRAS.476.5479P} are $\sim$0.25 dex larger than ours. This discrepancy is likely due to their use of relatively brighter 
galaxy samples ($i_{AB} < 25$), which may result in a lower number of smaller galaxies in the sample.
The steeper size evolution derived by \citet{Shibuya2019ApJ...871..164S} may be partly 
attributed to the inclusion of sources at $2 < z < 3$ in their fitting range. 
In this redshift interval, both the intrinsic evolution of the galaxies and 
the selection effect of \citet{Shibuya2019ApJ...871..164S} leads to larger sizes at lower redshifts, 
causing an apparent upturn in the size-redshift relation at low redshifts. 
When considering only their data within the redshift range of $3 < z < 7$, the median size is $\sim$ 0.5 kpc, 
which is about 0.2 dex larger than our measurement and consistent with the result 
of \citet{Paulino2018MNRAS.476.5479P}. 
Although \citet{Shibuya2019ApJ...871..164S} applied a luminosity cut ($L_{\mathrm{UV}} = (0.12\,\text{--}\, 1)L_{z=3}^{*}$) to exclude relatively faint sources at low redshift, this selection yields a nearly constant luminosity of $M_{UV} \sim -19.5$ across $3 < z < 7$. This may have excluded smaller, fainter galaxies 
at higher redshifts as well, thereby biasing their size measurements toward larger values.

\begin{figure*}[t!]
\centering
\includegraphics[width=1\textwidth]{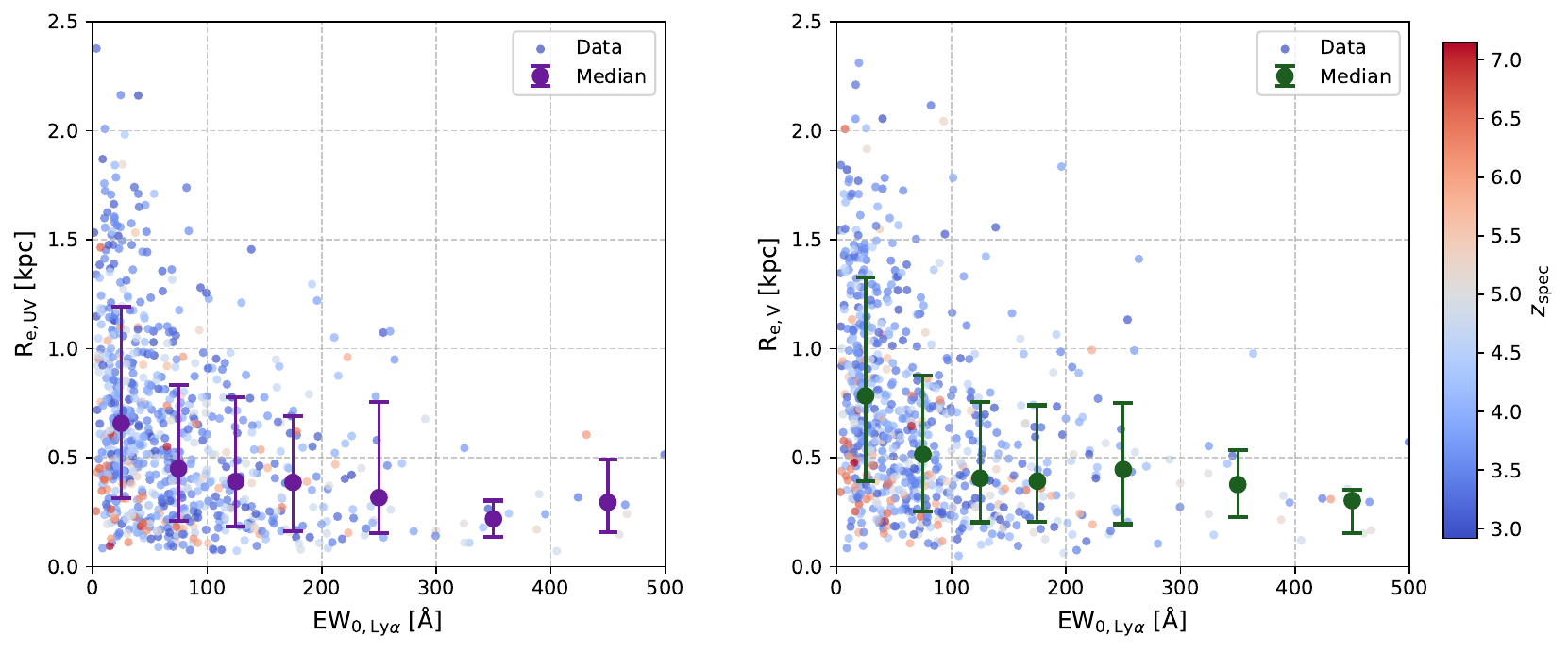} 
\caption{
The half-light radius of LAEs as a function of Ly$\alpha$ rest-frame equivalent width. 
The left panel shows the rest-frame UV radius, and the right panel shows the rest-frame optical radius. 
All data points are color-coded by spectroscopic redshift. 
The error bars represent the 16th to 84th percentile range.}

\label{fig:figure_6}
\end{figure*}

Furthermore, part of the discrepancies in the reported size evolution of LAEs among different studies 
may arise from differences in Ly$\alpha$ equivalent width (EW$_{0}$). 
As shown in Figure~\ref{fig:figure_6}, LAEs with higher EW$_{0}$ in our JWST sample 
are generally more compact, a trend consistent with earlier findings from 
HST-based studies \citep[e.g.,][]{Kerutt22}. This correlation suggests that sample selection 
effects -- particularly those introduced by photometric selection criteria -- 
could bias the interpretation of LAE size evolution.

\subsection{Comparison of the Size-Mass Relation with Previous Works} \label{sec:5.2}

As shown in Figure \ref{fig:size_mass}, the best-fit slopes of the size-mass relations 
for LAEs remain consistent ($\alpha \sim 0.20$) across both redshift intervals 
of $3 \lesssim z < 4$ and $4 < z < 7$. 
For comparison with the size-mass relations of typical SFGs, we overlay recent results 
based on JWST/NIRCam observations. To ensure a fair comparison with our LAEs, 
which are generally smaller in both size and stellar mass, we exclusively adopt the results 
from \citet{Allen2024arXiv241016354A}. This study examined SFGs at $3 < z < 9$ with stellar masses 
between $10^8$ and $10^{11}~M_{\odot}$ and specific star formation rate (sSFR) 
exceeding 0.2$t_{H}^{-1}$, using data from CEERS, PRIMER-UDS, and PRIMER-COSMOS surveys. 
The stellar masses were estimated based on a broken power-law IMF as described 
in \citet{Eldridge2017PASA...34...58E}, which is based on \citet{Kroupa1993MNRAS.262..545K}. 
For consistency, we apply the mass conversion prescribed 
by \citet{Madau2014ARA&A..52..415M} when plotting their results.

As shown in the left panel of Figure \ref{fig:size_mass}, within the redshift range $3 < z < 4$, 
the slope of the size-mass relation from \citet{Allen2024arXiv241016354A} ($\alpha \sim 0.20$) 
is broadly consistent with ours, though their sizes are on average larger by approximately 0.1 dex. 
This offset can be attributed to the intrinsically larger sizes of typical SFGs compared to LAEs 
in this redshift interval. 
In the right panel of Figure \ref{fig:size_mass}, the size-mass relation 
of our LAEs at $4 < z < 7$ closely matches that of \citet{Allen2024arXiv241016354A} 
in both slope ($\alpha \sim 0.23$) and size normalization over the subinterval $4 < z < 6$. 
It is noted that since \citet{Allen2024arXiv241016354A} provided separate 
relations for $4 < z < 5$ and $5 < z < 6$, 
we combined their results to allow a direct comparison with our broader redshift bin. 
Although our sample spans $4 < z < 7$, the small number of sources at $6 < z < 7$ 
has little influence on the overall trend, making the comparison with $4 < z < 6$ results 
of \citet{Allen2024arXiv241016354A} well justified. The slight deviation at the high-mass end 
is likely due to the limited number of massive LAEs in our sample.

\section{SUMMARY} \label{sec:6}

In this study, we investigate the size evolution of 876 spectroscopically confirmed LAEs 
at redshifts $3 \lesssim z < 7$, using JWST/NIRCam imaging data from the 
GOODS-S, COSMOS, and UDS fields. Our LAEs sample spans a $\rm Ly \alpha$ luminosity range 
of $\log(L_{\mathrm{Ly}\alpha}) \sim 40.3\,\text{--}\,43.35 \,\mathrm{erg\, s^{-1}}$ and 
an equivalent width range of $EW_{0, \mathrm{Ly}\alpha} \sim 3\,\text{--}\,1200 \,\text{\AA}$, 
representing a well-characterized sample with high completeness. 
In addition to tracing the size evolution, we compare the rest-frame 
optical and UV sizes of LAEs and present, for the first time, the optical size-mass relation 
for a large LAE sample. The main conclusions are summarized as follows:

\begin{enumerate}[itemsep=1em]
    \item The effective radius $R_{\rm e}$ of LAEs as a function of redshift follows $ R_{\rm e,UV} \propto (1 + z)^{-0.91 \pm 0.10} $ in the rest-frame UV-band and $ R_{\rm e,V} \propto (1 + z)^{-0.93 \pm 0.18} $ in the V-band 
within the range of $3 \lesssim z < 7$.
Specifically, in the rest-frame UV, the median size increases from $R_{\rm e} = 0.32_{-0.09}^{+0.11}$\,kpc
at $z \sim 7$ to $R_{\rm e} = 0.62_{-0.15}^{+0.18}$\,kpc at $z \sim 3$ as redshift decreases. 
In the rest-frame V band, the size increases from $R_{\rm e} = 0.36_{-0.16}^{+0.23}$\,kpc
at $z \sim 7$ to $R_{\rm e} = 0.71_{-0.30}^{+0.47}$\,kpc at $z \sim 3$.
\vspace{1em}
    \item The slopes of the optical size-mass relation for LAEs are consistent across two redshift intervals 
$3 \lesssim z < 4$ and $4 < z < 7$. The best-fit relations are $\log(R_{\rm e,V}) = (0.20 \pm 0.01) \times \log(M_/M_\odot) - (1.83 \pm 0.12)$ for $3 \lesssim z < 4$, and $\log(R_{\rm e,V}) = (0.20 \pm 0.03) \times \log(M_/M_\odot) - (1.91 \pm 0.23)$ for $4 < z < 7$. 
A comparative analysis of SED-derived stellar masses, with and without the inclusion of JWST MIRI/F770W 
photometry, reveals that MIRI data have only a minor impact on the mass estimates for LAEs at $z > 4$. 
When compared with recent studies of star-forming galaxies at similar redshifts, 
our results show a broadly consistent evolutionary trend in the size-mass relation. However, 
LAEs exhibit slightly smaller sizes at fixed stellar mass, particularly at lower redshifts.
\vspace{1em}
    \item The sizes of LAEs in the rest-frame UV and V bands are statistically consistent,
with a small median logarithmic offset of $\log(R_{\rm e,V}/R_{\rm e,UV}) \approx 0.03$.
This constancy implies that LAEs at $3 \lesssim z < 7$ have weak or negligible UV-optical color gradients.
\end{enumerate}

This study explores the size evolution of LAEs across multiple photometric bands, as well as the scaling relation between size and stellar mass. Leveraging spatially resolved, multi-wavelength diagnostics from combined JWST and HST observations, future analyses will enable increasingly precise constraints on the evolution of morphological structures, star formation histories, and dust attenuation laws in LAEs.

\section{acknowledgements}

We gratefully acknowledge the valuable feedback from the reviewer and the editorial team, which has significantly contributed to improving the quality of this work. This work is supported by the National Natural Science Foundation of China (NSFC  grants No. 12273052, 11733006, 12090040, 12090041, 12073051 and 12503015), the science research grants from  the China Manned Space Project (No.CMS-CSST-2021-A04). 
NL acknowledges the support from the Ministry of Science and Technology of China (No. 2020SKA0110100), the science research grants from the China Manned Space Project (No. CMS-CSST-2021-A01), and the CAS Project for Young Scientists in Basic Research (No. YSBR-062). 
This work utilizes observational data acquired through the NASA/ESA/CSA James Webb Space Telescope, retrieved from the Mikulski Archive for Space Telescopes (MAST) maintained by the Association of Universities for Research 
in Astronomy, located at the Space Telescope Science Institute. 
The data used in this study come from the following JWST programs: 1176 (PI Rogier Windhorst), 1180 (PI Daniel Eisenstein), 1210 (PI Nora Luetzgendorf), 1283 (PI Goeran Oestlin), 1286 (PI Nora Luetzgendorf), 1287 (PI Nora Luetzgendorf), 1837 (PI James Dunlop), 1895 (PI Pascal Oesch), 1963 (PIs Christina Williams, Michael Maseda, and Sandro Tacchella), 2079 (PIs Steven Finkelstein, Casey Papovich, and Norbert Pirzkal), 2198 (PIs Laia Barrufet and Pascal Oesch), 2514 (PIs Christina Williams and Pascal Oesch), 2516 (PIs Jacqueline Hodge and Elisabete da Cunha), 3215 (PIs Daniel Eisenstein and Roberto Maiolino), 3990 (PIs Takahiro Morishita, Charlotte Mason, Tommaso Treu, and Michele Trenti), 6541 (PI Eiichi Egami).

\software{\texttt{Astropy} \citep{Astropy2013A&A...558A..33A, Astropy2018AJ....156..123A, Astropy2022ApJ...935..167A}, \texttt{photutils} \citep{larry_bradley_2024_13989456}, \texttt{CIGALE} \citep{Boquien2019A&A...622A.103B, Yang2020MNRAS.491..740Y, Yang2022ApJ...927..192Y}, \texttt{Pysersic} \citep{pasha2023pysersic}, \texttt{PetroFit} \citep{geda2022petrofit}, \texttt{Statmorph} \citep{rodriguez2019optical}, \texttt{GalfitM} \citep{haussler2013megamorph, vika2013megamorph}}

\bibliography{ms.bib}
\bibliographystyle{aasjournal}

\appendix
\section{Representative Morphological Fitting Diagnostics for LAEs 
Using Multiple Software Packages}\label{size_fitting}

\begin{figure*}[h!]
\centering
\includegraphics[width=0.95\textwidth]{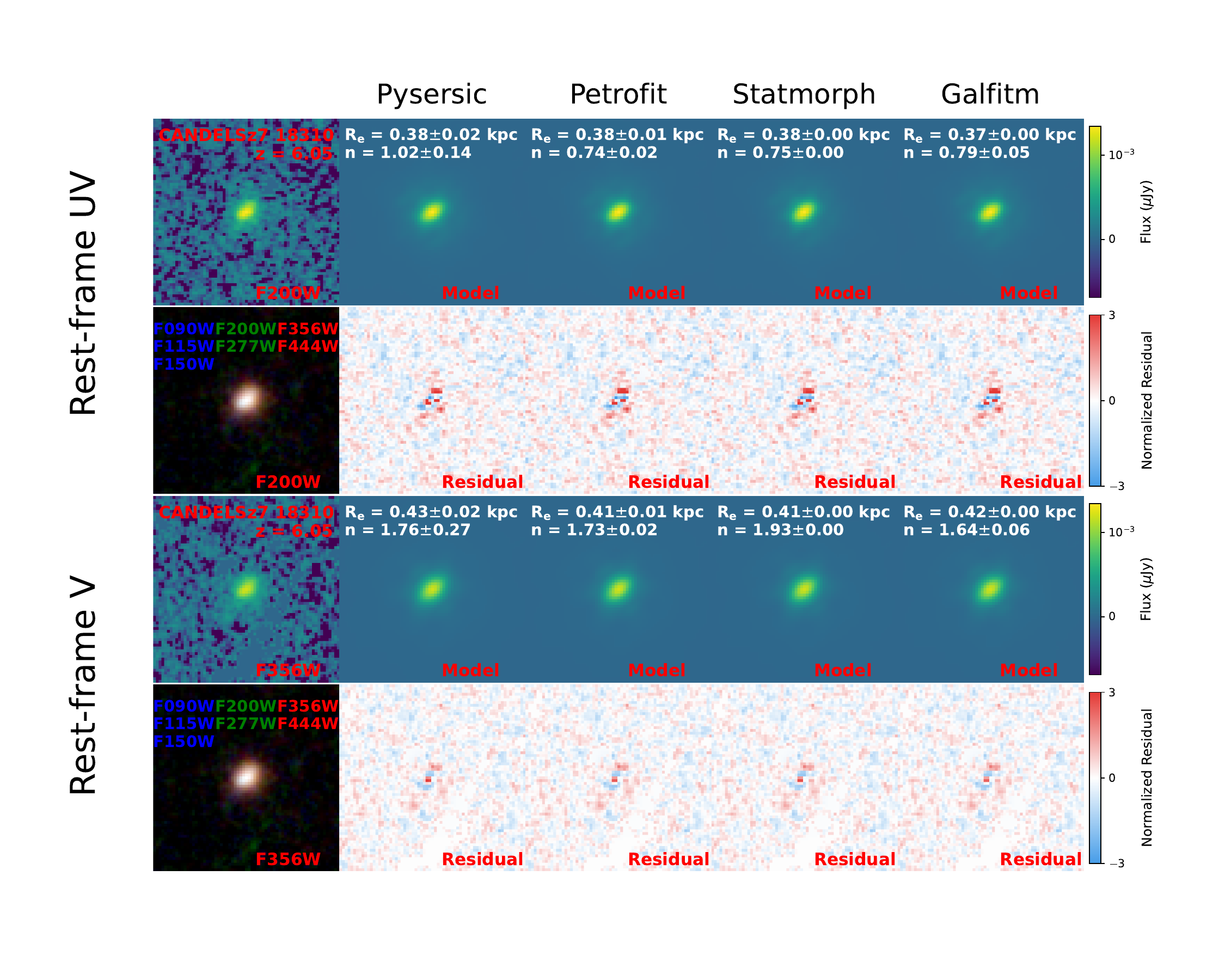} 
\caption{
Representative example of morphological fitting for an isolated LAE using 
four independent software packages. The figure displays the raw cutout images, model reconstructions, 
and corresponding residual maps. Pseudocolor images (combining multiple bands) are also 
included for visual context. Each cutout covers a $2''\times2''$ field of view.
}

\label{fig:figure_appendix1_1}
\end{figure*}

\begin{figure*}[h!]
\centering
\includegraphics[width=0.90\textwidth]{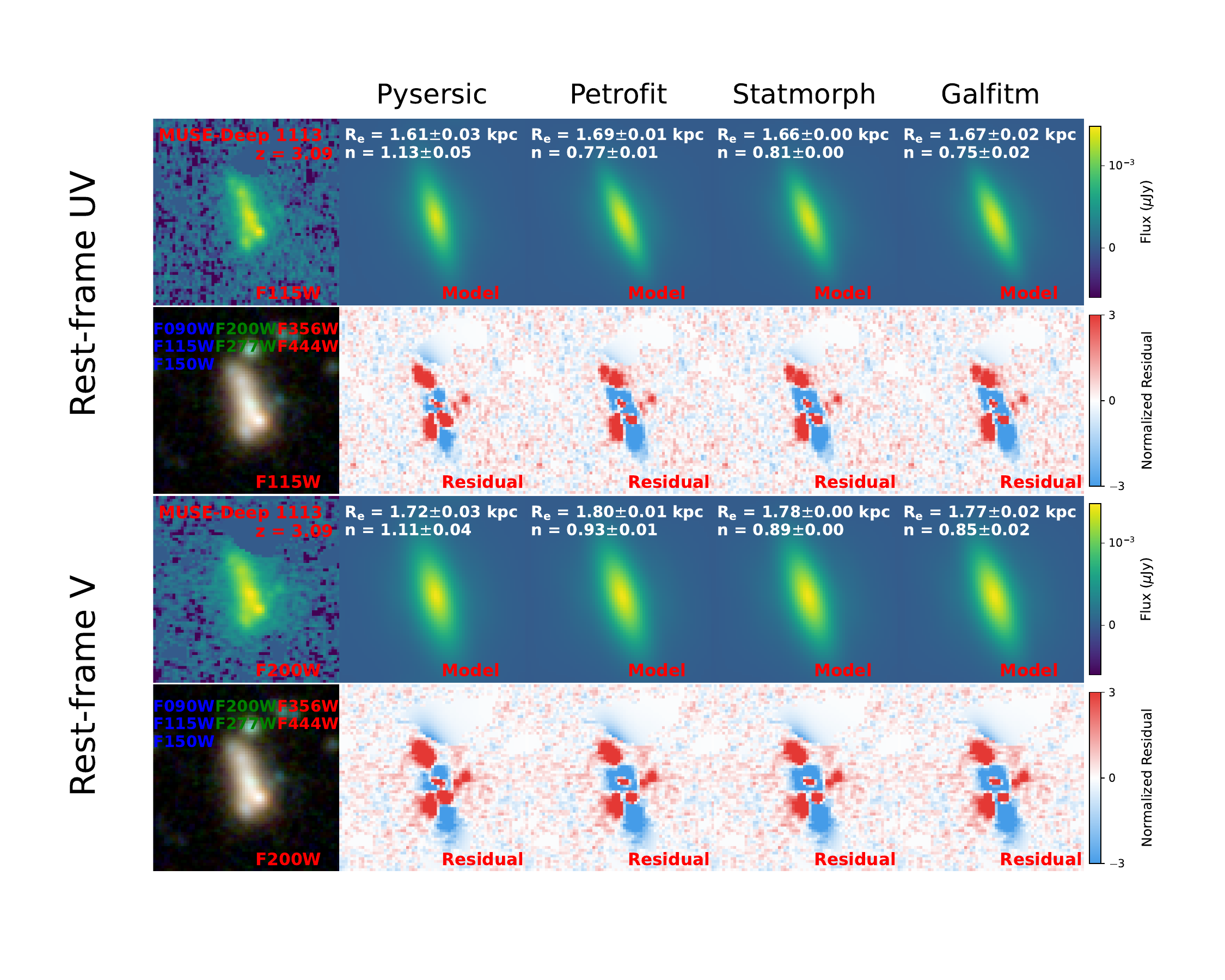} 
\caption{
Representative example of morphological fitting for a clumpy LAE 
using four independent software packages. 
The meanings of the imaging panels and symbols are the same as in Figure~\ref{fig:figure_appendix1_1}.
}

\label{fig:figure_appendix1_2}
\end{figure*}

\begin{figure*}[h!]
\centering
\includegraphics[width=0.90\textwidth]{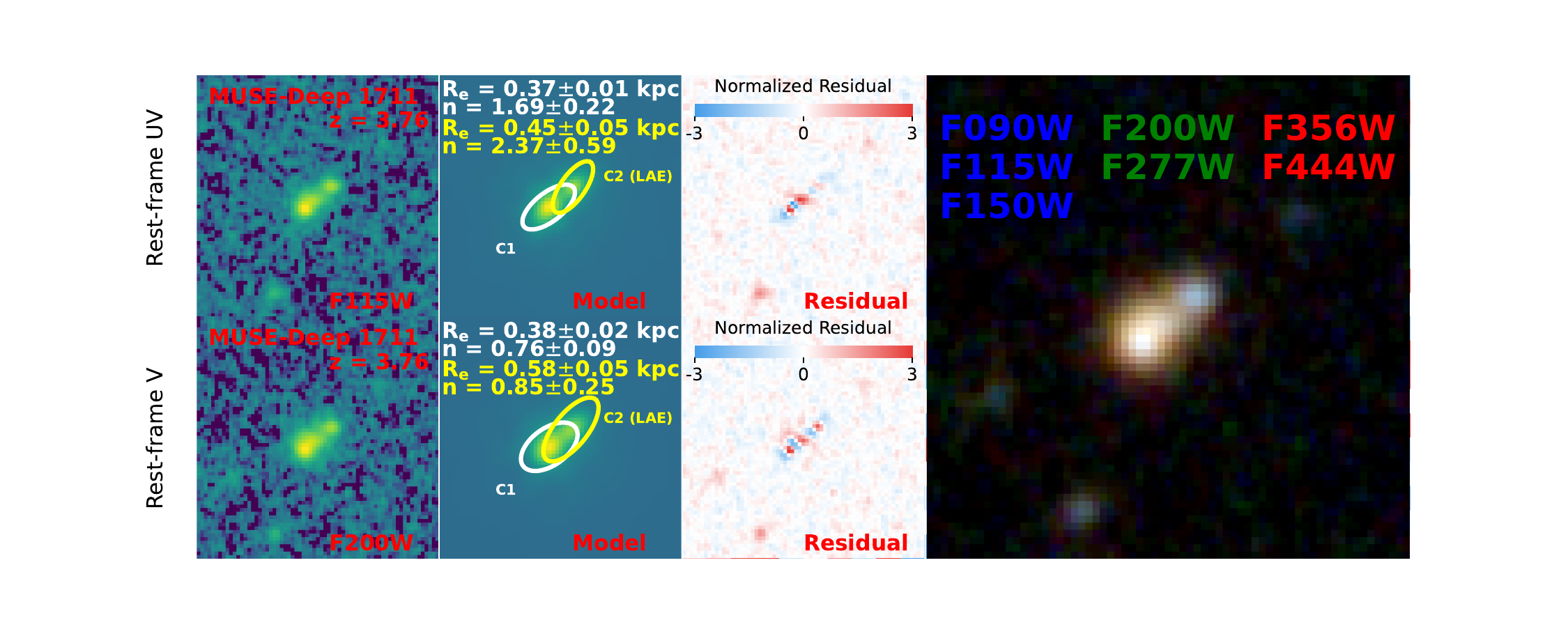} 
\caption{
Representative example of morphological fitting for a merging system using 
the \texttt{FitMulti} module in \texttt{PySersic}. 
The meanings of the imaging panels and symbols are the same as in Figure~\ref{fig:figure_appendix1_1}.
}

\label{fig:figure_appendix1_3}
\end{figure*}

Figures~\ref{fig:figure_appendix1_1}, \ref{fig:figure_appendix1_2} and \ref{fig:figure_appendix1_3} present representative fitting results for isolated sources, clumpy structures, and merging systems in our sample. 
For isolated sources, we performed multi-method validation using four independent software packages. 
Clumpy galaxies with poorly resolved or indistinct substructures are modeled as single-component systems.
Merging systems are fitted using the \texttt{FitMulti} module in \texttt{PySersic}. 
Each panel, covering a $2''\times2''$ field of view, displays rest-frame UV-band and optical 
V-band morphological analyses, enabling direct comparison across wavelengths. 
Detailed band selection criteria are provided in Section \ref{sec:3.1}.

\clearpage

\clearpage
\section{The Impact of MIRI/F770W Data on Stellar Mass Estimates: A Comparative Analysis}\label{stellar_mass}

\begin{figure*}[h!]
\centering
\includegraphics[width=1\textwidth]{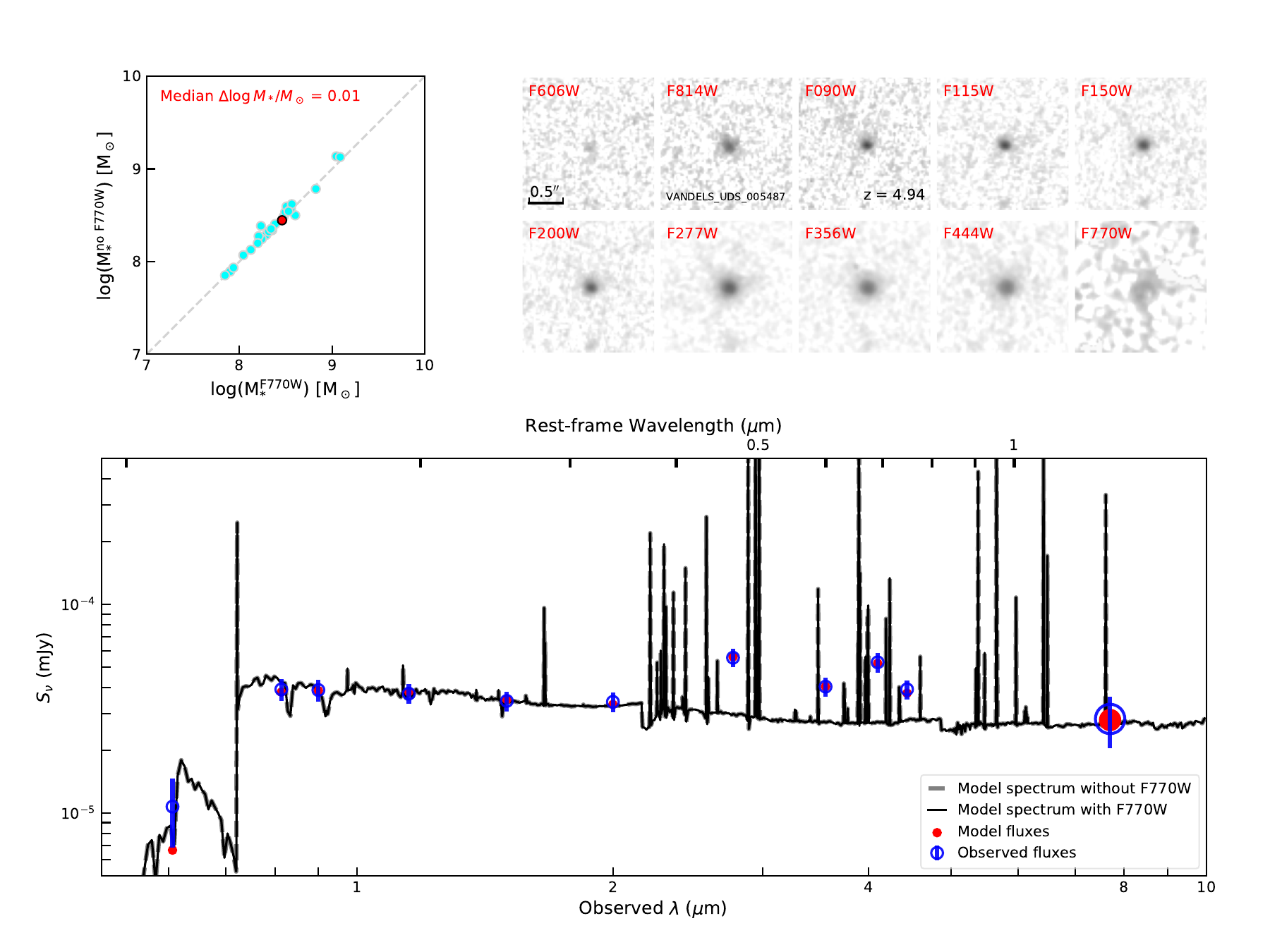} 
\caption{
Upper left: Comparison of stellar masses derived with and without the inclusion of MIRI/F770W photometry 
for LAEs in the COSMOS and UDS fields at 4 $<$ z $<$ 7. 
The median value of $\Delta(\log(M_{*}^{F770W}/M_\odot) - \log(M_{*}^{\mathrm{no}\ F770W}/M_\odot))$ 
is shown in the upper-left corner. Cyan symbols represent the stellar mass discrepancy 
$\Delta(\log(M_{*}^{F770W}/M_\odot) - \log(M_{*}^{\mathrm{no}\ F770W}/M_\odot))$, while 
the red dot marks the source displayed in the upper-right and bottom panels. 
Upper right: $2''\times2''$ JWST and HST cutout images of the source indicated by the red dot 
in the upper-left panel, corresponding to $z=4.94$. 
Bottom panel: SED of the source highlighted in red in the upper-left panel. 
Black solid lines show SED fits including MIRI/F770W photometry; 
gray dashed curves show fits excluding this band. 
The difference in best-fit stellar mass is negligible.
}
\label{fig:figure_appendix2}
\end{figure*}

As illustrated in Figure~\ref{fig:figure_appendix2}, 
we investigate the impact of longer-wavelength JWST/MIRI F770W data on stellar mass estimation 
by performing comparative SED fitting for LAEs in the COSMOS and UDS fields -- 
with and without F770W photometry. 
The results show that excluding F770W data leads to a negligible overestimation 
of stellar mass ($<0.1~dex$), indicating that the absence of JWST/MIRI observations 
introduces minimal systematic bias in stellar mass measurements for these galaxies.

\clearpage

\section{Basic properties of our LAE sample}\label{LAE_catalog}

This table lists the basic properties of our LAE sample, including coordinates, redshift, stellar mass, half-light radius, equivalent width, and more. The complete machine-readable version of this table is available online.

\begin{table}[ht]
\centering
\setlength{\tabcolsep}{3pt}
\caption{The physical properties of 876 LAEs}
\label{tab:table3}
\small
\begin{tabular}{lccccccccc}
\hline
ID & R.A. & Decl. & $z_{\rm spec}$ & EW$_{0}$ & log($L_{\rm Ly\alpha}$) & R$_{e,UV}$ & R$_{e,V}$ & $\log{M_\ast/M_\odot}$ & Survey \\
   & [deg] & [deg] &  & [\AA] & [erg/s] & [kpc] & [kpc] &  &  \\
\hline
VANDELS\_UDS\_386029 & 34.22146 & -5.09429 & 4.16 & 20.76$^{+5.43}_{-5.43}$ & 42.28$^{+0.07}_{-0.09}$ & 1.07$^{+0.00}_{-0.00}$ & 1.33$^{+0.02}_{-0.02}$ & 8.76$^{+0.12}_{-0.12}$ & VANDELS \\
23802 & 34.22835 & -5.14743 & 6.63 & 7.00$^{+0.00}_{-0.00}$ & 41.93$^{+0.00}_{-0.00}$ & 1.46$^{+0.00}_{-0.00}$ & 2.01$^{+0.00}_{-0.00}$ & 8.41$^{+0.13}_{-0.13}$ & CANDELSz7 \\
VANDELS\_UDS\_003586 & 34.23051 & -5.25616 & 4.83 & 29.65$^{+3.44}_{-3.44}$ & 42.54$^{+0.02}_{-0.02}$ & 0.22$^{+0.02}_{-0.02}$ & 0.53$^{+0.03}_{-0.03}$ & 8.56$^{+0.10}_{-0.10}$ & VANDELS \\
15559 & 34.23158 & -5.18979 & 6.04 & 79.00$^{+0.00}_{-0.00}$ & 42.51$^{+0.00}_{-0.00}$ & 0.96$^{+0.00}_{-0.00}$ & 0.86$^{+0.03}_{-0.03}$ & 8.26$^{+0.15}_{-0.15}$ & CANDELSz7 \\
VANDELS\_CDFS\_226868 & 53.01233 & -27.78404 & 3.69 & 50.44$^{+6.01}_{-6.01}$ & 42.39$^{+0.02}_{-0.02}$ & 0.98$^{+0.03}_{-0.03}$ & 1.10$^{+0.02}_{-0.02}$ & 9.18$^{+0.20}_{-0.20}$ & VANDELS \\
VANDELS\_CDFS\_225147 & 53.01429 & -27.79660 & 3.40 & 11.56$^{+3.20}_{-3.20}$ & 41.62$^{+0.08}_{-0.10}$ & 1.45$^{+0.00}_{-0.00}$ & 1.70$^{+0.05}_{-0.05}$ & 9.54$^{+0.15}_{-0.15}$ & VANDELS \\
15404 & 53.03388 & -27.77801 & 6.29 & 44.00$^{+0.00}_{-0.00}$ & 42.60$^{+0.00}_{-0.00}$ & 0.32$^{+0.04}_{-0.04}$ & 0.38$^{+0.02}_{-0.02}$ & 9.38$^{+0.16}_{-0.16}$ & CANDELSz7 \\
156044309 & 53.03540 & -27.78316 & 4.75 & 87.60$^{+19.32}_{-19.32}$ & 42.63$^{+0.46}_{-0.46}$ & 0.11$^{+0.01}_{-0.01}$ & 0.12$^{+0.01}_{-0.01}$ & 7.98$^{+0.16}_{-0.16}$ & MUSE\_wide \\
158026116 & 53.03699 & -27.80897 & 4.06 & 40.59$^{+12.80}_{-12.80}$ & 42.16$^{+0.66}_{-0.66}$ & 0.48$^{+0.07}_{-0.07}$ & 0.41$^{+0.01}_{-0.01}$ & 8.67$^{+0.17}_{-0.17}$ & MUSE\_wide \\
15178 & 53.05589 & -27.77956 & 6.27 & 48.00$^{+0.00}_{-0.00}$ & 42.60$^{+0.00}_{-0.00}$ & 0.19$^{+0.01}_{-0.01}$ & 0.40$^{+0.02}_{-0.02}$ & 9.04$^{+0.15}_{-0.15}$ & CANDELSz7 \\
1446 & 53.12550 & -27.78822 & 3.24 & 12.91$^{+1.79}_{-1.79}$ & 41.81$^{+0.04}_{-0.04}$ & 1.62$^{+0.09}_{-0.09}$ & 1.71$^{+0.04}_{-0.04}$ & 9.56$^{+0.11}_{-0.11}$ & MUSE\_Deep \\
7071 & 53.12736 & -27.78515 & 4.76 & 97.77$^{+13.13}_{-13.13}$ & 42.58$^{+0.01}_{-0.01}$ & 0.28$^{+0.01}_{-0.01}$ & 0.22$^{+0.00}_{-0.00}$ & 8.16$^{+0.21}_{-0.21}$ & MUSE\_Deep \\
215004022 & 150.08475 & 2.23333 & 3.24 & 159.12$^{+23.53}_{-23.53}$ & 42.73$^{+0.27}_{-0.27}$ & 0.47$^{+0.07}_{-0.07}$ & 0.25$^{+0.00}_{-0.00}$ & 6.91$^{+0.18}_{-0.18}$ & MUSE\_wide \\
208061363 & 150.08507 & 2.21337 & 5.02 & 107.18$^{+26.03}_{-26.03}$ & 42.83$^{+0.46}_{-0.46}$ & 0.51$^{+0.00}_{-0.00}$ & 0.49$^{+0.01}_{-0.01}$ & 7.60$^{+0.25}_{-0.25}$ & MUSE\_wide \\
7499 & 150.08916 & 2.26949 & 5.86 & 19.00$^{+0.00}_{-0.00}$ & 42.36$^{+0.00}_{-0.00}$ & 0.22$^{+0.03}_{-0.03}$ & 0.16$^{+0.00}_{-0.00}$ & 8.06$^{+0.12}_{-0.12}$ & CANDELSz7 \\
13679 & 150.09904 & 2.34363 & 7.15 & 15.00$^{+0.00}_{-0.00}$ & 42.74$^{+0.00}_{-0.00}$ & 0.09$^{+0.01}_{-0.01}$ & 0.48$^{+0.00}_{-0.00}$ & 8.74$^{+0.15}_{-0.15}$ & CANDELSz7 \\

...  & ...  &  ...  &  ...  &  ...  &  ...  &  ...  &  ...  &  ...  &  ... \\
\hline
\end{tabular}

\end{table}

\clearpage

\section{Representative Examples of Axis Ratio Measurements for LAEs}\label{axial ratio}

\begin{figure*}[h!]
\centering
\includegraphics[width=1\textwidth]{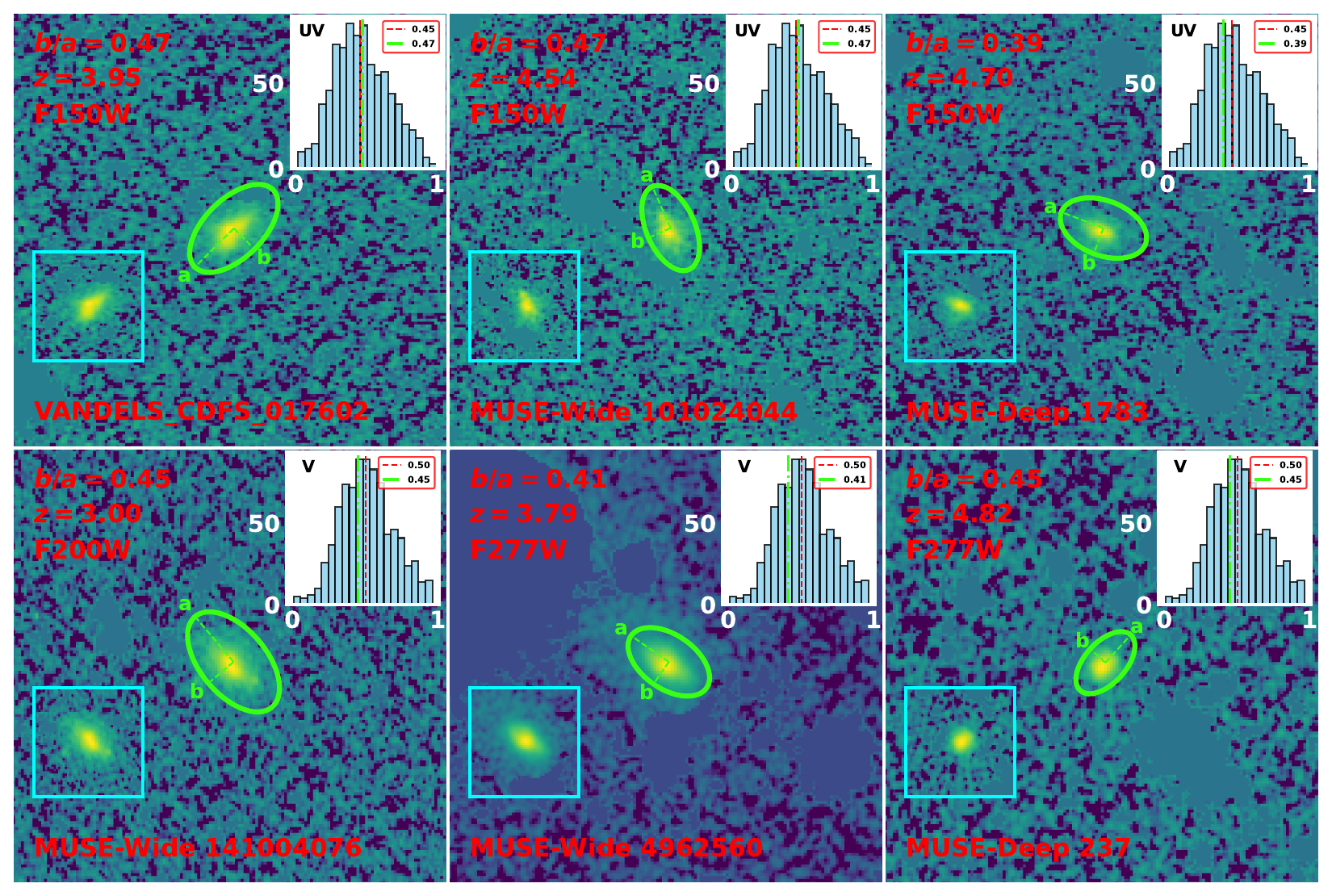} 
\caption{
Representative examples of axis ratio measurements for a subset of LAEs. 
The top three panels show measurements derived from rest-frame UV observations, 
while the bottom three panels display those from rest-frame V-band data. 
In each panel, a histogram of the overall axial ratio distribution for the full LAE sample 
is shown in the upper right corner. 
The red dashed line marks the median axial ratio of the sample, 
and the green dashed line indicates the axial ratio of the individual galaxy shown. 
The upper left corner lists the galaxy's axial ratio, redshift, and the observational band used. 
A zoomed-in view of the galaxy, along with its identifier, is presented in the lower left corner. 
Each main image spans $4.5''\times4.5''$, while the inset covers $2''\times2''$. 
The green ellipse represents the S\'{e}rsic model fitted to the source, 
with its semi-major and semi-minor axes indicated by green dashed lines.
}
\label{fig:figure_appendix3}
\end{figure*}

Figure~\ref{fig:figure_appendix3} presents the axial ratio measurements for a subset of LAEs, 
derived from both rest-frame UV and rest-frame V-band observations. The figure also shows 
the overall axial ratio distributions of the full sample. For our LAEs, the peaks of the axial ratio distributions 
lie in the range 0.45-0.50, in broad agreement with the results 
reported by \citet{Paulino2018MNRAS.476.5479P}.

\end{document}